\documentclass[10pt,english,twocolumn,twoside]{IEEEtran}
\usepackage[T1]{fontenc}
\usepackage{amsmath}
\usepackage{amssymb}
\usepackage{graphicx}

\makeatletter
\ifCLASSINFOpdf
\else
\fi
\usepackage[table]{xcolor}
\usepackage{rotating}
\usepackage{slashbox,pict2e}
\usepackage{cases}
\usepackage{enumitem}
\usepackage{footnote}
\makesavenoteenv{tabular}
\makesavenoteenv{table}

\usepackage{floatrow}
\usepackage[caption=false]{subfig}

\usepackage[algo2e, ruled, vlined, linesnumbered]{algorithm2e}

\def\hlinewd#1{%
\noalign{\ifnum0=`}\fi\hrule \@height #1 %
\futurelet\reserved@a\@xhline}
\newcolumntype{"}{@{\hskip\tabcolsep\vrule width 1pt\hskip\tabcolsep}}

\@ifundefined{showcaptionsetup}{}{%
 \PassOptionsToPackage{caption=false}{subfig}}
\usepackage{subfig}
\makeatother

\usepackage{babel}
\begin{document}
% merge modification from Tania
% merge modification from Prof. Tsui
% accept the changes
% include my modification
% merge modification from Prof. Tsui
% modify Fig. 6
% accept changes
% figure and table layout
% spell check
% latest modification from Prof. Tsui
% check the modification on the printed paper
% accept the changes from Prof. Tsui
% spell check
% check the reference list
% accept changes to the reference list
% final draft for 2nd round review% can use linebreaks \\ within to get better formatting as desired
% generate the latex source files for final submission

\title{A Low-Latency List Successive-Cancellation Decoding Implementation
for Polar Codes}

\author{YouZhe~Fan,~\IEEEmembership{Member,~IEEE,} ChenYang~Xia,~\IEEEmembership{Student
Member,~IEEE,} Ji~Chen,~\IEEEmembership{Student Member,~IEEE,}
Chi-ying~Tsui,~\IEEEmembership{Senior Member,~IEEE,} Jie~Jin,
Hui~Shen,~\IEEEmembership{Member,~IEEE,} and Bin~Li,~\IEEEmembership{Member,~IEEE}% <-this % stops a space
\thanks{This work has been published in part in the 40th International Conference
on Acoustics, Speech and Signal Processing (ICASSP'2015).

Y.-Z. Fan, C.-Y. Xia, J. Chen, and C.-Y. Tsui are with the Department
of Electronic and Computer Engineering, the Hong Kong University of
Science and Technology, Clear Water Bay, Kowloon, Hong Kong (e-mail:
\{jasonfan, cxia, jchenbh\}@connect.ust.hk, eetsui@ust.hk).

J. Jin, H. Shen, and B. Li are with the Communications Technology
Research Lab., Huawei Technologies, Shenzhen, P. R. China (e-mail:
\{steven.jinjie, henry.shenhui, binli.binli\}@huawei.com).%
}}

\maketitle


\markboth{Accepted for publication in IEEE Journal on Selected Areas in Communications}{Accepted for publication in IEEE Journal on Selected Areas in Communications}% The only time the second header will appear is for the odd numbered pages



%\IEEEpeerreviewmaketitle

%\boldmath

\begin{abstract}
Due to their provably capacity-achieving performance, polar codes
have attracted a lot of research interest recently. For a good error-correcting
performance, list successive-cancellation decoding (LSCD) with large
list size is used to decode polar codes. However, as the complexity
and delay of the list management operation rapidly increase with the
list size, the overall latency of LSCD becomes large and limits the
applicability of polar codes in high-throughput and latency-sensitive
applications. Therefore, in this work, the low-latency implementation
for LSCD with large list size is studied. Specifically, at the system
level, a selective expansion method is proposed such that some of
the reliable bits are not expanded to reduce the computation and latency.
At the algorithmic level, a double thresholding scheme is proposed
as a fast approximate-sorting method for the list management operation
to reduce the LSCD latency for large list size. A VLSI architecture
of the LSCD implementing the selective expansion and double thresholding
scheme is then developed, and implemented using a UMC 90 nm CMOS technology.
Experimental results show that, even for a large list size of 16,
the proposed LSCD achieves a decoding throughput of 460 Mbps at a
clock frequency of 658 MHz.
\end{abstract}


% \fi
%\IEEEpeerreviewmaketitle

\section{Introduction}

\IEEEPARstart{A}{s} the first family of error-correcting codes
provably achieving the channel capacity with explicit construction,
polar codes are a major breakthrough in coding theory \cite{Arikan}.
Due to their low encoding and decoding complexities, polar codes have
drawn a lot of research interest recently \cite{Urbanke_13}-\cite{BUPT_CM}.

\textit{Successive-cancellation decoding} (SCD) was proposed in \cite{Arikan}
for decoding polar codes. It was shown that SCD asymptotically achieves
the channel capacity when the code length $N$ is large \cite{Arikan}.
Moreover, the computational complexity of the SCD algorithm is low,
in the order of $N\log_{2}N$ \cite{Arikan}. Therefore, the SCD algorithm
and its hardware implementation have been extensively studied recently
\cite{SSC}-\cite{HW_Arikan}. However, for polar codes with short-to-medium
code length, the error-correcting performance of SCD is unsatisfactory.
For example, as shown in \cite{list}, compared with the low-density
parity-check (LDPC) code with similar code length and code rate, the
SNR penalty of SCD for $N=2048$ polar codes is greater than 1 dB
for a bit error rate of $10^{-5}$. Hence, to improve the performance
of polar codes with short-to-medium code length, SCDs generating multiple
codeword candidates were proposed. They are \textit{list successive-cancellation
decoding} (LSCD) \cite{list}, \cite{list_BUPT} and its variants
\cite{stack_BUPT}-\cite{SD_BUPT}. 

During the decoding of one codeword, LSCD generates $\mathcal{L}$
codeword candidates where $\mathcal{L}$ is called the \textit{list
size}. The value of $\mathcal{L}$ determines the trade-off between
the error-correcting performance and the computational complexity.
From \cite{list}, the LSCD approaches the \textit{maximum likelihood
decoding} (MLD) performance of polar codes with a moderate list size.
However, this performance is still not comparable with that of the
advanced error-correcting codes such as Turbo codes and LDPC codes.
To this end, to further improve the error-correcting performance,
cyclic redundancy check (CRC) code is serially concatenated with the
polar codes and the CRC bits are used to choose the valid codeword
from the candidates of the LSCD \cite{list}, \cite{CRC_BUPT}, \cite{CRC_Bin}.
With the help of the CRC code, the LSCD of polar codes achieves or
even exceeds the error-correcting performance of Turbo codes \cite{ICC_BUPT}
and LDPC codes \cite{list}. However, this performance improvement
is at the cost of a larger list size (e.g., $\mathcal{L}=16$ or 32)
and hence the complexity of the corresponding LSCD becomes high. The
high computational complexity also results in an LSCD architecture
with high decoding latency and low throughput.%
\footnote{Non-overlapped decoding architecture is assumed in this work; i.e.,
only one codeword is decoded each time in the hardware. Hence, the
higher the decoding latency is, the lower the decoding throughput
will be. Moreover, except otherwise stated, the latency in this work
is given in the number of clock cycles.%
} This limits the applicability of polar codes in high-throughput and
latency-sensitive applications. In this work, a low-latency LSCD architecture
is explored, aiming at promoting polar codes as a competitive coding
candidate in both the error-correcting and hardware implementation
aspects.

LSCD mainly consists of two classes of operations: 1) SCD operations
for generating each of the $\mathcal{L}$ codeword candidates, and
2) \textit{list management} (LM) operations for maintaining the $\mathcal{L}$
(locally) best codeword candidates in the list. SCD operations are
serial in nature and hence affect the decoding latency. LM operations
involve the finding of the best $\mathcal{L}$ out of $2\mathcal{L}$
candidates and maintaining the copy of the candidates. This requires
sorting and copying operations of which the complexity increases rapidly
with $\mathcal{L}$. To achieve a low latency, existing LSCD architectures
apply optimizations at either the algorithmic or architectural level.
As the first work on LSCD, \textit{lazy copy} was proposed in \cite{list}
to reduce the data copying complexity and hence the latency for the
LM operation. The corresponding gate-level implementation was detailed
in \cite{TCASII_EPFL}. In \cite{ICASSP_EPFL} and \cite{Asilomar_Parhi},
the operand of the SCD operation was changed from the \textit{log-likelihood}
(LL) value to the \textit{log-likelihood ratio} (LLR), resulting in
a simplified data path and improved clock frequency as well as a smaller
memory data storage. To reduce the latency introduced by the SCD operation,
multiple bits of a codeword were decoded at the same time in \cite{SiPS_Lehigh_Rate01}-\cite{TVLSI_Parhi}.
In \cite{ISCAS_Xiaohu}, the pre-computation look-ahead technique
was used to reduce the SCD latency by half, at the cost of a larger
memory. However, all these LSCD architectures \cite{TCASII_EPFL}-\cite{ISCAS_Xiaohu}
were designed for a small list size ($\mathcal{L}\leq4$).%
\footnote{In \cite{SiPS_Lehigh_Parallel}, an architecture for LM operation
supporting $\mathcal{L}=8$ was proposed; however, the overall LSCD
architecture was not presented. An LSCD algorithm for list size up
to $\mathcal{L}=128$ was discussed in \cite{SiPS_Gross}. However,
it was implemented on the PC platform instead of in VLSI.%
} With the increase of the list size, both the computational complexity
and the logic delay of the LM operation become larger. Therefore,
to support LSCD for $\mathcal{L}=8$ with a reasonable clock frequency,
up to three pipeline stages were inserted in the LM operation and
three cycles were needed for each LM operation in \cite{ISCAS_Lehigh}.
This resulted in a long decoding latency. In \cite{sTSP_EPFL}, the
serial sorting operation in the LM operation was parallelized at the
architectural level \cite{Parallel_Sorting}, and the latency of the
resulting LSCD architecture was reduced for $\mathcal{L}=8$. However,
as shown in \cite{ISCAS_EPFL}, even using a parallel architecture,
the logic delay of the LM operation keeps increasing with the list
size, and it deteriorates the clock frequency of the overall LSCD
architecture for a larger list size ($\mathcal{L}>8$). Therefore,
in this work, we concentrate on reducing the latency introduced by
LM operations, especially for a large list size $\mathcal{L}$. 
\begin{figure}
\centering{}\includegraphics{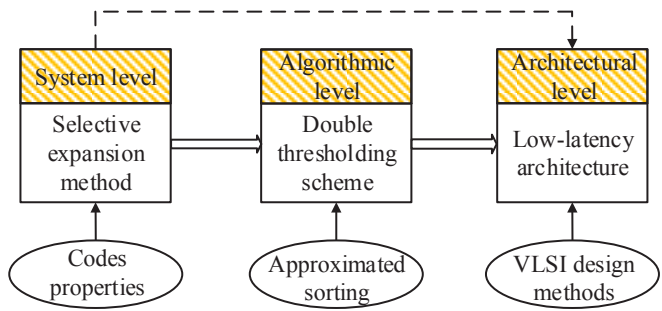}\caption{Low-latency LSCD design flowgraph.}
\end{figure}

This work achieves low-latency LSCD implementation by performing optimizations
at the system, algorithmic, and architectural level, as depicted in
Fig. 1. At the system level, a method called \textit{selective expansion}
(SE) is proposed based on the properties of polar codes. From \cite{Arikan},
each \textit{source word} bit of the polar code's codeword corresponds
to a \textit{synthetic channel}, and different synthetic channels
have different reliabilities. In the SE method, only those bits associated
with the less reliable synthetic channels are decoded with the LSCD,
while the more reliable bits are decoded by the SCD \cite{patent_bin}.
As a result, the LM operation (and its associated latency) for the
reliable bits are not needed. To implement the SE method on the LSCD
architecture, an optimization problem is formulated to determine which
bits are decoded by the LSCD, such that the latency saving is maximized
for a given error-correcting performance requirement of the system.
We note that, similar to SE, a concurrent work \cite{IET_UNSW} was
proposed to reduce the complexity of the LSCD by utilizing the synthetic
channel characteristics. However, the methodology and the goal of
this work and ours are different. At the algorithmic level, an approximated
LM operation called the \textit{double thresholding scheme} (DTS)
is proposed. Instead of exactly maintaining the $\mathcal{L}$ (locally)
best codeword candidates, the DTS keeps the $\mathcal{L}$ almost-the-best
codeword candidates in the list such that the performance degradation
introduced is negligible \cite{icassp}. Compared with the original
LM operation, the DTS is parallel in nature and its logic delay is
independent of the list size. Hence, the latency of the LM operation
is not increased, even for a large list size. Finally, at the architectural
level, an efficient LSCD architecture based on the DTS is proposed.
By optimizing the schedule and logic of the blocks related to the
LM operation, a low-latency LSCD implementation is achieved, even
for $\mathcal{L}=16$.

The remainder of this paper is organized as follows. The construction
of polar codes and the algorithm of LSCD are reviewed in Section II.
Section III presents the proposed SE method for reducing the latency
of LSCD. The DTS is detailed in Section IV and Section V presents
the LSCD architecture with a low decoding latency. In Section VI,
the simulation results of the error-correcting performance of the
proposed low-latency LSCD architecture are presented. The ASIC implementation
results of the proposed architecture are also shown. Finally, Section
VII concludes the work.

\section{Preliminaries}

In this section, the \textit{channel polarization phenomenon} \cite{Arikan}
discovered by Arıkan is firstly reviewed, and it is fundamental to
the SE method discussed in Section III. After that, the construction
of polar codes and the algorithm of LSCD are reviewed.

\subsection{Channel Polarization Phenomenon}

Consider a \textit{binary-input discrete memoryless channel}, denoted
as $W$: $\mathcal{X}\rightarrow\mathcal{Y}$, with an input alphabet
$\mathcal{X}\in\left\{ 0,1\right\} $ and an output alphabet $\mathcal{Y}$.
Channel $W$ is specified by the channel transition probabilities
$W\left(y|x\right)$ with $x\in\mathcal{X}$ and $y\in\mathcal{Y}$.

Let $W_{N}$: $\mathcal{X}^{N}\rightarrow\mathcal{Y}^{N}$ denote
$N$ independent copies of channel $W$, where $N=2^{n}$ and $n\in\mathbb{N}$.
Channel $W_{N}$ can be described by the channel transition probabilities
and is given by
\begin{equation}
W_{N}\left(\mathbf{y}_{N}|\mathbf{x}_{N}\right)=\prod_{i=0}^{N-1}W\left(y_{i}|x_{i}\right),\label{eq:parallel_channel}
\end{equation}
where $\mathbf{x}_{N}\in\mathcal{X}^{N}$ and $\mathbf{y}_{N}\in\mathcal{Y}^{N}$
are the input and the output of $W_{N}$, respectively.

Let $\mathbf{u}_{N}\in\mathcal{X}^{N}$ be a binary vector one-to-one
mapped to $\mathbf{x}_{N}$ by the following relation:
\begin{equation}
\mathbf{x}_{N}^{T}=\mathbf{u}_{N}^{T}\mathbf{F}^{\otimes n},\label{eq:encoding}
\end{equation}
where $\mathbf{x}^{T}$ is the transpose of $\mathbf{x}$, and $\mathbf{F}^{\otimes n}$
is the $n^{\textrm{th}}$ \textit{Kronecker power} of the kernel matrix
$\mathbf{F}$. Since $\mathbf{F}\triangleq\left[\begin{array}{cc}
1 & 0\\
1 & 1
\end{array}\right]$, $\mathbf{F}^{\otimes n}$ is invertible.

Based on \eqref{eq:encoding}, $N$ \textit{synthetic channels} are
obtained from $W_{N}$. They are denoted as $W_{N}^{i}$: $\mathcal{X}\rightarrow\mathcal{X}^{i}\times\mathcal{Y}^{N}$,
where $i\in\left\{ 0,1,\ldots,N-1\right\} $. The transition probabilities
of channel $W_{N}^{i}$ are given by \begin{equation} W_{N}^{i}\!\left(\mathbf{y}_{N},\mathbf{u}_{0}^{i-1}|u_{i}\right)=\!\sum_{\mathbf{u}_{i+1}^{N-1}\in\mathcal{X}^{N-i-1}}\!\frac{1}{2^{N-1}}W_{N}\!\left(\mathbf{y}_{N}|\mathbf{x}_{N}\right),\label{eq:synthetic_channel} \end{equation}where
$\mathbf{x}_{N}$ and $\mathbf{u}_{N}$ are related by \eqref{eq:encoding}.
$\mathbf{x}_{a}^{b}$ denotes the sub-vector of $\mathbf{x}$ with
a starting and ending index of $a$ and $b$. From \eqref{eq:synthetic_channel},
the input of a synthetic channel $W_{N}^{i}$ is a binary bit $u_{i}\in\mathcal{X}$,
and its output includes the $W_{N}$ output $\mathbf{y}_{N}$ and
the side information of the $i$ preceding bits $\mathbf{u}_{0}^{i-1}$.
To evaluate the performance of the synthetic channels, a probability
of error $P_{e}\left(i\right)$ is associated with each channel $W_{N}^{i}$.
Under \textit{maximum likelihood decoding} (MLD), $P_{e}\left(i\right)$
is given as \begin{equation} P_{e}\!\left(i\right)\!=\!\sum_{\mathbf{u}_{0}^{i-1}\!,\mathbf{y}_{N}}\!\frac{\min\!\left\{\! W_{N}^{i}\!\left(\mathbf{y}_{N}\!,\!\mathbf{u}_{0}^{i-1}|0\right)\!,\!W_{N}^{i}\!\left(\mathbf{y}_{N}\!,\!\mathbf{u}_{0}^{i-1}|1\right)\!\right\} }{2}\!,\label{eq:Pe} \end{equation}where
$\mathbf{u}_{0}^{i-1}\in\mathcal{X}^{i}$ and $\mathbf{y}_{N}\in\mathcal{Y}^{N}$,
and $u_{i}$ assumes the value of $\mathcal{X}$ with equal probability.
For any given $N$, the values of the $P_{e}\left(i\right)$s can
be found efficiently by the density evolution techniques, as presented
in \cite{ISIT_Mori}-\cite{Vardy}.

Arıkan's \textit{Channel Polarization Theorem} studies the behavior
of the synthetic channel $W_{N}^{i}$ \cite{Arikan}. One key observation
of the theorem is that when $N\rightarrow\infty$, the performance
of the synthetic channel $W_{N}^{i}$ is polarized; i.e., except for
a vanishing fraction of $W_{N}^{i}$s, the rest of the $W_{N}^{i}$s
are either almost noise-free ($P_{e}\left(i\right)\rightarrow0$)
or almost useless ($P_{e}\left(i\right)\rightarrow0.5$). For a finite
value of $N$, the $P_{e}\left(i\right)$s of the synthetic channels
are getting close to either 0 or 0.5, and the $P_{e}\left(i\right)$s
are different for different $W_{N}^{i}$s \cite{ISIT_Mori}-\cite{Vardy}.

\subsection{Construction of Polar Codes}

Based on the channel polarization phenomenon, the construction of
polar codes is simple. In a polar coding scheme, \eqref{eq:encoding}
represents the encoding operation of a length $N$ polar code. Vectors
$\mathbf{u}_{N}$ and $\mathbf{x}_{N}$ are the \textit{source word}
and \textit{codeword}, respectively. A \textit{rate} $R=K/N$ polar
code is specified by the \textit{frozen set} $\mathcal{A}^{c}\subset\left\{ 0,1,\ldots,N-1\right\} $
of cardinality $\left|\mathcal{A}^{c}\right|=N-K$ and the \textit{information
set} $\mathcal{A}$ defined as $\mathcal{A}=\left\{ 0,1,\ldots,N-1\right\} \setminus\mathcal{A}^{c}$.
The $K$ source word bits $u_{i}$ ($i\in\mathcal{A}$) deliver the
\textit{information bits}, and the remaining $N-K$ bits $u_{i}$
($i\in\mathcal{A}^{c}$) are the \textit{frozen bits}. Since the frozen
bits are set to a value, e.g. 0, known to both the encoder and the
decoder, the block-error probability $P_{b}$ of polar codes is bounded
by \cite{Arikan},
\begin{equation}
P_{b}\leq\sum_{i\in\mathcal{A}}P_{e}\left(i\right).\label{eq:block_error_rate}
\end{equation}

From \eqref{eq:block_error_rate}, choosing the $K$ indices with
the smallest $P_{e}\left(i\right)$s in $\mathcal{A}$ minimizes the
block-error probability $P_{b}$. From the discussion in Section II-A,
if $K$ is not greater than the number of the almost noise-free synthetic
channels, a reliable communication is achieved by the polar codes.

If $r$-bit CRC code is used in polar codes, to maintain a fixed code
rate $R$, the information set $\mathcal{A}$ is extended such that
$\left|\mathcal{A}\right|=NR+r$ by switching $r$ most reliable frozen
bits to the information bits. These extended bits deliver the CRC
code bits of the original $NR$ information bits. In the LSCD, only
the codeword candidate passing the CRC check is output as the decoding
result. 
\begin{figure}
\centering{}\includegraphics{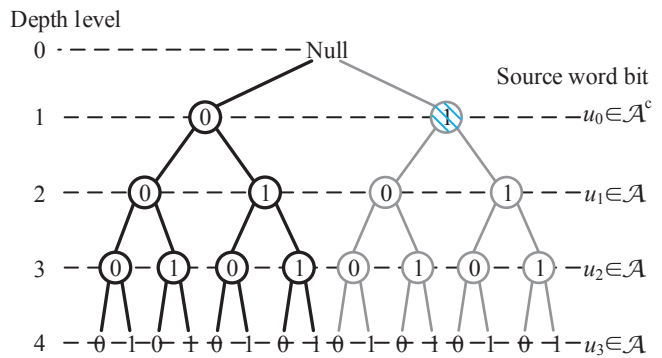}\caption{Decoding tree for an $N=4$ polar code.}
\end{figure}

\subsection{List Successive-Cancellation Decoding}

The decoding process of polar codes can be treated as a search problem
in the \textit{decoding tree}. As an example, Fig. 2 shows the decoding
tree for an $N=4$ polar code. In general, the decoding tree of a
length-$N$ polar code is a depth-$N$ binary tree, with $u_{i}$
mapped to the nodes at depth $i+1$. As shown in Fig. 2, its root
node represents a null state, and the left and right children at depth
$i+1$ represent $u_{i}=0$ and $u_{i}=1$, respectively. Therefore,
a path from the root node to a depth-$i$ node represents a sub-vector
$\mathbf{u}_{0}^{i-1}\in\mathcal{X}^{i}$, and it is called a \textit{decoding
path}. Specifically, a \textit{complete decoding path} is a path from
the root node to the leaf node that represents a vector $\mathbf{u}_{N}\in\mathcal{X}^{N}$.
The value of each bit of $\mathbf{u}_{N}$ is shown in the corresponding
node lying at this decoding path. If $u_{i}$ is a frozen bit, it
only assumes a preset value, e.g. 0. Consequently, the right-hand
sub-tree rooted at the depth-$\left(i+1\right)$ node is pruned, as
$\mathbf{u}_{N}$s included in this sub-tree are not valid source
words. For example, if $\mathcal{A}^{c}=\left\{ 0\right\} $, the
gray sub-tree in Fig. 2 is pruned. As a result, each complete decoding
path in the pruned decoding tree is one-to-one corresponding to a
valid source word of the polar code, denoted as $\mathcal{U}=\left\{ \mathbf{u}_{N}|u_{i}\left(i\in\mathcal{A}^{c}\right)=0\right\} $.
In the subsequent discussion, let $\mathbf{u}_{N}\in\mathcal{U}$
be the transmitted source word, and the task of the decoder is to
find a complete decoding path $\hat{\mathbf{u}}_{N}\in\mathcal{U}$
to decode $\mathbf{u}_{N}$.

The MLD of polar codes exhaustively searches all the complete decoding
paths in the decoding tree and generates the \textit{likelihood} $\Pr\left(\mathbf{y}_{N}|\hat{\mathbf{u}}_{N}\right)$
for each complete decoding path $\hat{\mathbf{u}}_{N}\in\mathcal{U}$,
where 
\begin{equation}
\Pr\left(\mathbf{y}_{N}|\hat{\mathbf{u}}_{N}\right)=W_{N}\left(\mathbf{y}_{N}|\hat{\mathbf{u}}_{N}^{T}\mathbf{F}^{\otimes n}\right).\label{eq:likelihood}
\end{equation}
The decoding path $\hat{\mathbf{u}}_{N}^{\textrm{MLD}}$ with the
maximum $\Pr\left(\mathbf{y}_{N}|\hat{\mathbf{u}}_{N}\right)$ is
output as the decoding result.

To ease the implementation, likelihood $\Pr\left(\mathbf{y}_{N}|\hat{\mathbf{u}}_{N}\right)$
is represented by the \textit{path metric} $\gamma^{N}\left(\hat{\mathbf{u}}_{N}\right)$,
which is given by \cite{ICASSP_EPFL}:
\begin{equation}
\gamma^{N}\left(\hat{\mathbf{u}}_{N}\right)=-\log\left[\Pr\left(\mathbf{y}_{N}|\hat{\mathbf{u}}_{N}\right)\right]-\log\frac{\Pr\left(\hat{\mathbf{u}}_{N}\right)}{\Pr\left(\mathbf{y}_{N}\right)}.\label{eq:pm}
\end{equation}
For a given channel observation $\mathbf{y}_{N}$, the second term
in \eqref{eq:pm} is the same for all the source word $\hat{\mathbf{u}}_{N}$s.
Therefore, the MLD of polar codes is described by
\begin{equation}
\hat{\mathbf{u}}_{N}^{\textrm{MLD}}=\arg\min_{\hat{\mathbf{u}}_{N}\in\mathcal{U}}\gamma^{N}\left(\hat{\mathbf{u}}_{N}\right).\label{eq:MLD}
\end{equation}

Recently, \cite{ICASSP_EPFL} and \cite{Asilomar_Parhi} showed that
the path metric $\gamma^{N}\left(\hat{\mathbf{u}}_{N}\right)$ can
be expressed as
\begin{equation}
\gamma^{N}\left(\hat{\mathbf{u}}_{N}\right)=\sum_{i=0}^{N-1}\log\left\{ 1+\exp\left[\left(2\hat{u}_{i}-1\right)\cdot\Lambda^{i}\left(\hat{\mathbf{u}}_{0}^{i-1}\right)\right]\right\} ,\label{eq:pm_expansion}
\end{equation}
where $\hat{u}_{i}$ is the $i^{\textrm{th}}$ bit of the decoding
path $\hat{\mathbf{u}}_{N}$. $\Lambda^{i}\left(\hat{\mathbf{u}}_{0}^{i-1}\right)$
denotes the \textit{output LLR} of the synthetic channel $W_{N}^{i}$,
which is given as
\begin{equation}
\Lambda^{i}\left(\hat{\mathbf{u}}_{0}^{i-1}\right)=\log\frac{W_{N}^{i}\left(\mathbf{y}_{N},\hat{\mathbf{u}}_{0}^{i-1}|u_{i}=0\right)}{W_{N}^{i}\left(\mathbf{y}_{N},\hat{\mathbf{u}}_{0}^{i-1}|u_{i}=1\right)}.\label{eq:output_LLR}
\end{equation}
From \eqref{eq:output_LLR}, the value of $\Lambda^{i}\left(\hat{\mathbf{u}}_{0}^{i-1}\right)$
depends on the previous decoding path $\hat{\mathbf{u}}_{0}^{i-1}$,
and therefore each decoding path $\hat{\mathbf{u}}_{0}^{i-1}$ corresponds
to a different output LLR $\Lambda^{i}\left(\hat{\mathbf{u}}_{0}^{i-1}\right)$.
Using the alternative form of the path metric expressed in \eqref{eq:pm_expansion}
enables the use of LLR-based SCD in LSCD which leads to a lower logic
delay and memory requirement over its LL-based counterpart \cite{ICASSP_EPFL}-\cite{Asilomar_Parhi}.
\begin{figure}
\begin{centering}
\includegraphics{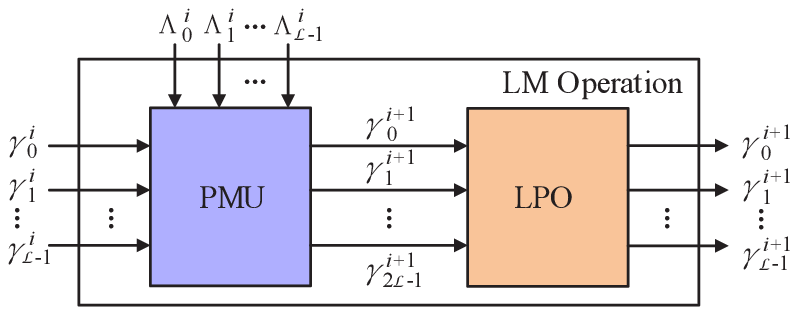}
\par\end{centering}

\caption{List management (LM) operation.}
\end{figure}

Similarly, a \textit{path metric $\gamma^{i}\left(\hat{\mathbf{u}}_{0}^{i-1}\right)$}
is associated with the decoding path $\hat{\mathbf{u}}_{0}^{i-1}$,
and is given as \begin{equation} \gamma^{i}\!\left(\hat{\mathbf{u}}_{0}^{i-1}\right)=\sum_{j=0}^{i-1}\log\!\left\{ 1+\exp\!\left[\left(2\hat{u}_{j}\!-\!1\right)\!\cdot\!\Lambda^{j}\!\left(\hat{\mathbf{u}}_{0}^{j-1}\right)\right]\right\} \!,\label{eq:pm_def} \end{equation}
where $\gamma^{0}=0$. Considering all the decoding path $\hat{\mathbf{u}}_{0}^{i-1}$s
at a certain depth of the decoding tree, the path metric $\gamma^{i}\left(\hat{\mathbf{u}}_{0}^{i-1}\right)$
and the output LLR $\Lambda^{i}\left(\hat{\mathbf{u}}_{0}^{i-1}\right)$
of each path is available. When the decoding path is extended to the
next depth, the path metric of $\hat{\mathbf{u}}_{0}^{i}$ is updated
as \begin{equation} \gamma^{i+1}\!\left(\!\hat{\mathbf{u}}_{0}^{i}\!\right)\!=\!\gamma^{i}\!\left(\!\hat{\mathbf{u}}_{0}^{i-1}\!\right)+\log\!\left\{ 1\!+\!\exp\!\left[\!\left(2\hat{u}_{i}\!-\!1\right)\!\cdot\!\Lambda^{i}\!\left(\!\hat{\mathbf{u}}_{0}^{i-1}\!\right)\!\right]\!\right\} \!,\label{eq:pm_update} \end{equation}where
the decoding path $\hat{\mathbf{u}}_{0}^{i}$ is extended from $\hat{\mathbf{u}}_{0}^{i-1}$.
Here, $\hat{u}_{i}$ can be either 0 or 1 if $i\in\mathcal{A}$. Otherwise,
$\hat{u}_{i}=0$. The operation in \eqref{eq:pm_update} is called
the \textit{Path Metric Update} (PMU) in this work. With the PMU,
the path metrics of all the paths $\hat{\mathbf{u}}_{N}$ are generated
and \eqref{eq:MLD} can be executed accordingly. Therefore, the MLD
can be regarded as a breadth-first search in the decoding tree.

Since there are $2^{K}$ complete decoding paths in the pruned decoding
tree, the MLD complexity is as large as $\mathcal{O}\left(2^{K}\right)$.
To achieve a reasonable decoding complexity, LSCD is proposed to obtain
a decoding performance close to that of MLD with a much smaller complexity.
For an LSCD with a list size of $\mathcal{L}$, at most $\mathcal{L}$
decoding paths are maintained at each depth of the decoding tree.
Therefore, after decoding $\log_{2}\mathcal{L}$ information bit $u_{i}$s,%
\footnote{To ease the discussion, $\mathcal{L}$ is assumed to be an integer
power of 2. The methodology of this work does not have a constraint
on the value of $\mathcal{L}$. %
} the decoding list has $\mathcal{L}$ decoding paths. In the subsequent
decoding, if $u_{i}$ is a frozen bit, $\mathcal{L}$ decoding paths
$\hat{\mathbf{u}}_{0}^{i}$s are extended from $\hat{\mathbf{u}}_{0}^{i-1}$s.
On the other hand, if $u_{i}$ is an information bit, $2\mathcal{L}$
decoding paths $\hat{\mathbf{u}}_{0}^{i}$s are extended from $\mathcal{L}$
$\hat{\mathbf{u}}_{0}^{i-1}$s. As a result, to maintain the list
size, a \textit{List Pruning Operation} (LPO) has to be executed.
Out of the $2\mathcal{L}$ decoding paths, the LPO keeps the $\mathcal{L}$
paths with the minimum path metrics and drops the rest. For simplicity,
the decoding path $\hat{\mathbf{u}}_{0}^{i-1}$ in the path metric
notation $\gamma^{i}\left(\hat{\mathbf{u}}_{0}^{i-1}\right)$ and
output LLR notation $\Lambda^{i}\left(\hat{\mathbf{u}}_{0}^{i-1}\right)$
are dropped in the subsequent discussion, and the $\mathcal{L}$ path
metrics (and output LLRs) are indexed by the subscript $l=0,1,\ldots,\mathcal{L}-1$.
In this work, as depicted in Fig. 3, the LPO together with the PMU
is denoted as the LM operation. 
\begin{figure}
\begin{centering}
\includegraphics{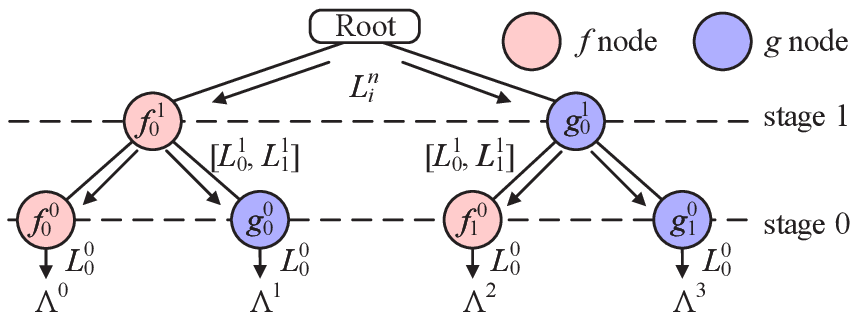}
\par\end{centering}

\caption{Scheduling tree for an $N=4$ polar code.}
\end{figure}

In the PMU operation specified by \eqref{eq:pm_update}, the output
LLR $\Lambda^{i}$ of each decoding path $\hat{\mathbf{u}}_{0}^{i-1}$
is required and it is generated by the SCD. The SCD operation for
a length-$N$ polar code can be represented by a depth-$n$ balanced
binary tree, called the \textit{scheduling tree} \cite{psn}. Fig.
4 shows an example of the scheduling tree for an $N=4$ polar code.
Its root node provides the \textit{input LLR} $L_{i}^{n}$s from the
channel observation $\mathbf{y}_{N}$ as follows:
\begin{equation}
L_{i}^{n}=\log W\left(y_{i}|x_{i}=0\right)-\log W\left(y_{i}|x_{i}=1\right),\label{eq:input_llr}
\end{equation}
where $i=0,1,\ldots,N-1$. The non-root nodes in the scheduling tree
are categorized into two types: the $f$ \textit{node} at the left-hand
child and the $g$ \textit{node} at the right-hand child. The $f$
node at stage $t$ executes the following $f$ \textit{function},
\begin{equation}
L_{j}^{t}=2\tanh^{-1}\left[\tanh\left(L_{j}^{t+1}/2\right)\tanh\left(L_{j+2^{t}}^{t+1}/2\right)\right],\label{eq:f_function}
\end{equation}
and the $g$ node executes the following $g$ \textit{function},
\begin{equation}
L_{j}^{t}=L_{j+2^{t}}^{t+1}+\left(-1\right)^{s_{j}}L_{j}^{t+1},\label{eq:g_function}
\end{equation}
where $j=0,1,\ldots,2^{t}-1$ and $L_{j}^{t}$s are the output LLRs
at stage $t$. From \eqref{eq:f_function} and \eqref{eq:g_function},
each function of the SCD has two LLRs as inputs and one LLR as output.
A node at stage $t$ of the scheduling tree includes $2^{t}$ functions,
and they can be executed in parallel. As a result, $2^{t}$ $L_{j}^{t}$s
are output by a node at stage $t$, and they are the inputs of its
two children in the next stage. 

The variable $s_{j}$ in \eqref{eq:g_function} is known as the \textit{partial-sum}
in \cite{semi_parallel_Gross} and \cite{psn}. The partial-sum $\mathbf{s}=\left[s_{0},s_{1},\ldots,s_{2^{t}-1}\right]$
is calculated from the previous decoding path $\tilde{\mathbf{u}}=\hat{\mathbf{u}}_{i-2^{t}}^{i-1}$
by 
\begin{equation}
\mathbf{s}^{T}=\tilde{\mathbf{u}}^{T}\mathbf{F}^{\otimes t}.\label{eq:partial_sum}
\end{equation}
Due to the data dependency introduced by the partial-sum, the decoding
schedule of the SCD follows the depth-first traversal of the scheduling
tree. As shown in Fig. 4, the $i^{\textrm{th}}$ leaf node of the
scheduling tree outputs the LLR of the synthetic channel $W_{N}^{i}$
as $\Lambda^{i}=L_{0}^{0}$ and hence $\Lambda^{i}$s are serially
generated. Based on $\Lambda^{i}$, if $i\in\mathcal{A}$, the MLD
of $u_{i}$ is given by
\begin{equation}
\Theta\left(\Lambda^{i}\right)=\begin{cases}
\begin{array}{l}
0\\
1
\end{array} & \begin{array}{l}
\textrm{if}\;\Lambda^{i}\geq0,\\
\textrm{else},
\end{array}\end{cases}\label{eq:hard_decision}
\end{equation}
where $\Theta\left(\Lambda^{i}\right)$ is the hard-decision function
based on the value of $\Lambda^{i}$. The probability of error for
$\hat{u}_{i}=\Theta\left(\Lambda^{i}\left(\mathbf{u}_{0}^{i-1}\right)\right)$,
i.e., $\Pr\left(\hat{u}_{i}\neq u_{i}\right)$, is given by $P_{e}\left(i\right)$
in \eqref{eq:Pe}. If $i\in\mathcal{A}^{c}$, $u_{i}$ is decoded
as 0.\begin{algorithm2e}[t] 
\caption{Procedure of LSCD}
$L\leftarrow 1$\tcp*[r]{initialize the actual size of the candidate list}
\For {$i=0,1,\cdots N-1$}{
\For (\tcp*[f]{given the $l^{\textrm{th}}$ decoding path $\hat{\mathbf{u}}_{0}^{i-1}$}) {$l=0,1,\cdots L-1$}{
update $\Lambda_{l}^{i}$ with \eqref{eq:f_function}-\eqref{eq:partial_sum}\tcp*[r]{SCD operation}
\eIf {$i\in\mathcal{A}^{c}$} {
extend $\hat{\mathbf{u}}_{0}^{i-1}$ to $\hat{\mathbf{u}}_{0}^{i}$ with $\hat{u}_{i}\leftarrow 0$\;
update $\gamma_{l}^{i+1}$ from $\gamma_{l}^{i}$ and $\Lambda_{l}^{i}$ with $\hat{u}_{i}\leftarrow 0$\tcp*[r]{PMU in \eqref{eq:pm_update}}
}{
extend $\hat{\mathbf{u}}_{0}^{i-1}$ to two path $\hat{\mathbf{u}}_{0}^{i}$s with $\hat{u}_{i}\leftarrow 0/1$\;
update $\gamma_{l}^{i+1}$s from $\gamma_{l}^{i}$ and $\Lambda_{l}^{i}$ with $\hat{u}_{i}\!\in\!\left\{ 0,1\right\}$\tcp*[r]{PMU in \eqref{eq:pm_update}}
$L\leftarrow L+1$;
}
}
\If (\tcp*[f]{LPO}) {$L>\mathcal{L}$}{
find $\mathcal{L}$ smallest $\gamma_{l}^{i+1}$s and corresponding $\hat{\mathbf{u}}_{0}^{i}$s\;
$L\leftarrow \mathcal{L}$;
}
}
\Return {$\hat{\mathbf{u}}_{N}$} passed the CRC check;
\end{algorithm2e}

Algorithm 1 summarizes the procedure of an LSCD with list size $\mathcal{L}$.
Line 4 indicates that the LSCD consists of $\mathcal{L}$ SCDs. They
are executed in parallel till a leaf node of the scheduling tree is
reached. With $\mathcal{L}$ output LLR $\Lambda_{l}^{i}$s, the decoding
path $\hat{\mathbf{u}}_{0}^{i-1}$s are extended to the next depth
of the decoding tree and the path metrics are updated by the PMU.
If the number of extended paths is greater than $\mathcal{L}$, the
LPO is executed. Note that the SCD operation has to be stalled till
the LPO is finished because the subsequent SCD operation needs the
knowledge of the previous path $\hat{\mathbf{u}}_{0}^{i}$, as discussed
in \eqref{eq:partial_sum}. As a result, the decoding schedule of
the LSCD can also be represented by the depth-first traversal of the
scheduling tree, except that the LM operation (Lines 5-14 in Algorithm
1) has to be executed at each leaf node of the scheduling tree. Hence,
the decoding latency of the LSCD depends on the latency of both the
SCD and LM operations.

Finally, it is noted that the PMU in \eqref{eq:pm_update} and the
$f$ function in \eqref{eq:f_function} are non-linear functions.
To simplify the hardware implementation, the PMU is approximated as
follows \cite{ICASSP_EPFL}-\cite{Asilomar_Parhi}, \cite{sTSP_EPFL}:
\begin{equation}
\begin{cases}
\begin{array}{l}
\gamma_{2l}^{i+1}=\gamma_{l}^{i}\\
\gamma_{2l+1}^{i+1}=\gamma_{l}^{i}+\left|\Lambda_{l}^{i}\right|
\end{array} & \begin{array}{l}
\textrm{if}\;\hat{u}_{i}=\Theta\left(\Lambda_{l}^{i}\right),\\
\textrm{if}\;\hat{u}_{i}=\overline{\Theta\left(\Lambda_{l}^{i}\right)},
\end{array}\end{cases}\label{eq:pmu_approx}
\end{equation}
where $l=0,1,\ldots,\mathcal{L}-1$, and $\gamma_{2l}^{i+1}$ and
$\gamma_{2l+1}^{i+1}$ denote the path metrics of the two path extensions
from the $l^{\textrm{th}}$ decoding path $\hat{\mathbf{u}}_{0}^{i-1}$,
respectively. Here, $\overline{x}$ is the complement of the binary
variable $x$. Similarly, the $f$ function is usually approximated
as \cite{semi_parallel_Gross}-\cite{HW_Arikan} \begin{equation} L_{j}^{t}=\textrm{sgn}\!\left(L_{j}^{t+1}\right)\oplus\textrm{sgn}\!\left(L_{j+2^{t}}^{t+1}\right)\min\!\left(\left|L_{j}^{t+1}\right|,\left|L_{j+2^{t}}^{t+1}\right|\right)\!,\label{eq:f_approx} \end{equation}where
$\textrm{sgn}\left(\cdot\right)$ and $\left|\cdot\right|$ represent
the sign bit and the magnitude of a variable, respectively. As hardware
implementation is discussed in this work, \eqref{eq:pmu_approx} and
\eqref{eq:f_approx} will be used for the corresponding calculation
except otherwise stated.

\section{Selective Expansion}

\subsection{Selective Expansion Scheme}

From the discussion in Section II-C, additional latency is introduced
by an LM operation, when $\mathcal{L}$ decoding paths are expanded
into $2\mathcal{L}$ paths for an information bit $u_{i}$ ($i\in\mathcal{A}$)
in the LSCD. In this section, we present a \textit{selective expansion}
(SE) scheme where the path expansion for some of the information bits
is not executed; i.e., $\mathcal{L}$ decoding paths are only extended
into $\mathcal{L}$ paths for those bits. As a result, the \textit{list
pruning operation} (LPO) is not needed and the associated latency
will not be added to the overall latency.

When an information bit $u_{i}$ ($i\in\mathcal{A}$) is decoded,
there are $\mathcal{L}$ surviving decoding paths $\hat{\mathbf{u}}_{0}^{i-1}$s
available due to the decoding of the previous $i$ bits. Assuming
that ultimately the LSCD will correctly decode the source word, there
exists one path $\mathbf{u}_{0}^{i-1}$ out of the $\mathcal{L}$
surviving decoding paths $\hat{\mathbf{u}}_{0}^{i-1}$s that will
lead to the correct decoding of the source word $\mathbf{u}_{N}$.
Consider the path extensions from $\mathbf{u}_{0}^{i-1}$. From \eqref{eq:output_LLR},
the output LLR of $\mathbf{u}_{0}^{i-1}$ is $\Lambda^{i}\left(\mathbf{u}_{0}^{i-1}\right)=\log W_{N}^{i}\left(\mathbf{y}_{N},\mathbf{u}_{0}^{i-1}|u_{i}=0\right)-\log W_{N}^{i}\left(\mathbf{y}_{N},\mathbf{u}_{0}^{i-1}|u_{i}=1\right)$.
From the discussion of \eqref{eq:hard_decision}, $\hat{u}_{i}$ assumes
either $\Theta\left(\Lambda^{i}\right)$ or $\overline{\Theta\left(\Lambda^{i}\right)},$
and the probability of error for $\hat{u}_{i}=\Theta\left(\Lambda^{i}\right)$
is $P_{e}\left(i\right)$. Therefore, if the decoding path $\mathbf{u}_{0}^{i-1}$
is only extended into a single path taking $\hat{u}_{i}$ as $\Theta\left(\Lambda^{i}\right)$,
the probability of this path extension leading to an incorrect decoding
of the transmitted source word $\mathbf{u}_{N}$ is then $P_{e}\left(i\right)$.
From the discussion in Section II-A, even inside the information set,
different bits have different $P_{e}\left(i\right)$s. If $u_{i}$
corresponds to a very reliable channel with a very low $P_{e}\left(i\right)$,
the probability of $\mathbf{u}_{0}^{i}$ not being in the candidate
list by only extending the path into a single path assuming $\hat{u}_{i}=\Theta\left(\Lambda^{i}\right)$
is small and the performance degradation introduced is negligible. 

Based on the above discussion, the SE method is proposed. It divides
the information set $\mathcal{A}$ into two subsets: the \textit{reliable
set} and the \textit{unreliable set}, denoted by $\mathcal{A}_{r}$
and $\mathcal{A}_{u}$, respectively. Only for those bits inside $\mathcal{A}_{u}$
are the $\mathcal{L}$ decoding paths expanded into $2\mathcal{L}$
paths. If $u_{i}$ is in $\mathcal{A}_{r}$, each of the $\mathcal{L}$
decoding paths is extended into a single path by taking $\hat{u}_{i}=\Theta\left(\Lambda^{i}\right)$.
Consequently, the LPO and the associated latency are saved for those
bits inside $\mathcal{A}_{r}$. Moreover, from \eqref{eq:pmu_approx},
since $\hat{u}_{i}$ is taken to be $\Theta\left(\Lambda^{i}\right)$,
no PMU operation is required. Next, the method of determining the
set $\mathcal{A}_{r}$ is discussed.

\subsection{Reliable Set for Selective Expansion}

To determine the reliable set $\mathcal{A}_{r}$ (or equivalently
$\mathcal{A}_{u}=\mathcal{A}\setminus\mathcal{A}_{r}$), the performance
of LSCD using the SE method is firstly analyzed. Let $\mathcal{M}_{\textrm{LSCD}}$
and $\mathcal{M}_{\textrm{SE}}$ denote the candidate lists output
from the conventional LSCD and the LSCD using the SE method, respectively.
We are mainly interested in the \textit{block-error event} that the
transmitted source word $\mathbf{u}_{N}$ is not in $\mathcal{M}_{\textrm{LSCD}}$
or $\mathcal{M}_{\textrm{SE}}$. The block-error event of the SE method
$\mathcal{E}_{\textrm{SE}}\triangleq\left\{ \mathbf{u}_{N}\notin\mathcal{M}_{\textrm{SE}}\right\} $
is given by \setlength{\arraycolsep}{0.1em} 
\begin{eqnarray} 
\mathcal{E}_{\textrm{SE}} & = & \mathcal{E}_{\textrm{SE}}^{\textrm{LSCD}}\cup\mathcal{E}_{\textrm{SE}}^{\overline{\textrm{LSCD}}}\label{eq:error_event_SE}\\
 & = & \left\{\! \mathbf{u}_{N}\!\notin\!\mathcal{M}_{\textrm{SE}},\!\mathbf{u}_{N}\!\notin\!\mathcal{M}_{\textrm{LSCD}}\! \right\} \cup\left\{ \!\mathbf{u}_{N}\!\notin\!\mathcal{M}_{\textrm{SE}},\!\mathbf{u}_{N}\!\in\!\mathcal{M}_{\textrm{LSCD}}\! \right\} \!,\nonumber  \end{eqnarray}
\setlength{\arraycolsep}{5pt}where $\mathcal{E}_{\textrm{SE}}^{\overline{\textrm{LSCD}}}$ and
$\mathcal{E}_{\textrm{SE}}^{\textrm{LSCD}}$ denote the error events
in the SE method that can and cannot be correctly decoded by the conventional
LSCD, respectively. In other words, $\mathcal{E}_{\textrm{SE}}^{\overline{\textrm{LSCD}}}$
is the error events introduced by the SE method, since otherwise they
can be decoded by the conventional LSCD. In addition, let $\mathcal{E}_{\textrm{LSCD}}\triangleq\left\{ \mathbf{u}_{N}\notin\mathcal{M}_{\textrm{LSCD}}\right\} $
be the block-error event of the conventional LSCD and we have $\mathcal{E}_{\textrm{SE}}^{\textrm{LSCD}}\subset\mathcal{E}_{\textrm{LSCD}}$.
A block-error event in $\mathcal{E}_{\textrm{SE}}^{\overline{\textrm{LSCD}}}$
occurs when we decode an information bit $u_{i}$, where $i\in\mathcal{A}_{r}$,
and the resulting $\mathcal{L}$ candidate paths do not include the
correct path $\mathbf{u}_{0}^{i}$. This event is denoted by $\mathcal{B}=\left\{ u_{i}\neq\Theta\left(\Lambda^{i}\left(\mathbf{u}_{0}^{i-1}\right)\right)|i\in\mathcal{A}_{r}\right\} $,
where $u_{i}$ is the $i^{\textrm{th}}$ bit of the transmitted source
word $\mathbf{u}_{N}$ and $\Lambda^{i}\left(\mathbf{u}_{0}^{i-1}\right)$
is obtained based on $\mathbf{u}_{0}^{i-1}$. Similar to the union
bound of the SCD in \eqref{eq:block_error_rate}, the probability
of event $\mathcal{B}$ satisfies
\begin{equation}
\Pr\left(\mathcal{B}\right)\leq\sum_{i\in\mathcal{A}_{r}}P_{e}\left(i\right).\label{eq:pe_removing}
\end{equation}
From the above discussion, the block-error probability of the LSCD
using the SE method, i.e., $P_{b}^{\textrm{SE}}\triangleq\Pr\left(\mathcal{E}_{\textrm{SE}}\right)$,
is upper bounded by
\begin{eqnarray}
P_{b}^{\textrm{SE}} & = & \Pr\left(\mathcal{E}_{\textrm{SE}}^{\textrm{LSCD}}\right)+\Pr\left(\mathcal{E}_{\textrm{SE}}^{\overline{\textrm{LSCD}}}\right)\nonumber \\
 & \leq & P_{b}^{\textrm{LSCD}}+\left(1-P_{b}^{\textrm{LSCD}}\right)\cdot\Pr\left(\mathcal{B}\right)\label{eq:union_bound_tmp_SE}\\
 & \leq & P_{b}^{\textrm{LSCD}}+\sum_{i\in\mathcal{A}_{r}}P_{e}\left(i\right),\nonumber 
\end{eqnarray}
where $P_{b}^{\textrm{LSCD}}\triangleq\Pr\left(\mathcal{E}_{\textrm{LSCD}}\right)$
denotes the block-error probability of the conventional LSCD.%
\footnote{It is assumed in this work that the value of $P_{b}^{\textrm{LSCD}}$
is already available and it can be obtained from the simulation. We
leave the theoretical analysis of $P_{b}^{\textrm{LSCD}}$ to our
future works.%
} Furthermore, to simplify the calculation of \eqref{eq:union_bound_tmp_SE},
$P_{e}\left(i\right)$s are approximated by the error probability
$P_{e}^{d}\left(i\right)$s of their degraded channels \cite{Vardy},
where $P_{e}\left(i\right)\leq P_{e}^{d}\left(i\right)$. As a result,
the upper bound on $P_{b}^{\textrm{SE}}$ is given by
\begin{equation}
P_{b}^{\textrm{SE}}\leq P_{b}^{\textrm{LSCD}}+\sum_{i\in\mathcal{A}_{r}}P_{e}^{d}\left(i\right).\label{eq:union_bound_SE}
\end{equation}
Based on \eqref{eq:union_bound_SE}, we define the upper bound of
the block-error probability degradation $\eta$ introduced by the
SE method as
\begin{equation}
\eta\left(\mathcal{A}_{r}\right)\triangleq\frac{\sum_{i\in\mathcal{A}_{r}}P_{e}^{d}\left(i\right)}{P_{b}^{\textrm{LSCD}}},\label{eq:BLER_upper_bound}
\end{equation}
and the block-error probability of the LSCD using SE is no greater
than $\left(1+\eta\right)P_{b}^{\textrm{LSCD}}$.

From the above performance analysis result, we formulate an optimization
problem given a constraint on the tolerable error-correcting performance
degradation $\epsilon$ as follows:
\begin{equation}
\begin{array}{ll}
\textrm{maximize} & \left|\mathcal{A}_{r}\right|\\
\textrm{subject to} & \mathcal{A}_{r}\subset\mathcal{A}\\
 & \eta\leq\epsilon.
\end{array}\label{eq:optimization_prob}
\end{equation}
The solution of \eqref{eq:optimization_prob} is the optimal set of
$\mathcal{A}_{r}$, as the objective function $\left|\mathcal{A}_{r}\right|$,
reflecting the latency saving achieved by the SE method, is maximized.

The optimal solution to problem \eqref{eq:optimization_prob} can
be obtained by sorting the information set $\mathcal{A}$ by $P_{e}^{d}\left(i\right)$
($i\in\mathcal{A}$) in ascending order and taking the first $k$
elements in the sorted $\mathcal{A}$ such that the corresponding
$\eta$ of this $k$-element set is just smaller than $\epsilon$.
For an information set $\mathcal{A}$ of polar codes with a given
$\epsilon$, the reliable set $\mathcal{A}_{r}$ of \eqref{eq:optimization_prob}
can be found offline accordingly.

\section{Double Thresholding Scheme}

For the SE method, the LM operation still has to be executed for those
unreliable information bits. The LPO needs to find the smallest $\mathcal{L}$
path metrics from the $2\mathcal{L}$ candidate inputs and sorting
method is required. However, the sorting operation introduces a large
latency, particularly when the list size is large. To reduce the latency,
parallel sorting can be used, but the computation complexity will
be very high for a large list size. Therefore, a low complexity sorting
operation is needed. In this section, a \textit{Double Thresholding
Scheme} (DTS) is proposed at the algorithmic level as a good approximation
of the conventional sorting method. Low complexity parallel comparisons
are executed in the DTS to find the surviving paths, and the latency
of the LPO is greatly reduced for a large list size $\mathcal{L}$.

\subsection{Properties of the Path Metric}

From Section II-C, the inputs to the LPO of bit $u_{i}$ are $2\mathcal{L}$
path metrics $\gamma_{k}^{i+1}$ ($k=0,1,\ldots,2\mathcal{L}-1$)
generated from the PMU as stated in \eqref{eq:pmu_approx}. To approximate
the LPO, the properties of the input path metrics are first studied.
Specifically, we are interested in the number of the path metrics
that are smaller than a certain value $T$, i.e., the cardinality
of the set $\Omega\left(T\right)$ which is defined as
\begin{equation}
\Omega\left(T\right)\triangleq\left\{ \gamma_{k}^{i+1}|\gamma_{k}^{i+1}<T\right\} .\label{eq:omiga}
\end{equation}
The properties related to the cardinality $\left|\Omega\left(T\right)\right|$
are stated as follows.\newtheorem{prop}{Proposition}\begin{prop}Assume
the $\mathcal{L}$ path metrics $\gamma_{l}^{i}$ ($l=0,1,\ldots,\mathcal{L}-1$)
input to the PMU are sorted and
\begin{equation}
\gamma_{0}^{i}<\gamma_{1}^{i}<\cdots<\gamma_{l}^{i}<\gamma_{l+1}^{i}<\cdots<\gamma_{\mathcal{L}-1}^{i}.\label{eq:sort_assumption}
\end{equation}
The cardinality of $\Omega\left(T\right)$, when $T=\gamma_{l}^{i}$,
satisfies
\begin{equation}
l\leq\left|\Omega\left(\gamma_{l}^{i}\right)\right|\leq2l.\label{eq:set_size}
\end{equation}
\end{prop}\begin{IEEEproof}[Proof]From \eqref{eq:pmu_approx} and
\eqref{eq:sort_assumption}, $\gamma_{0}^{i+1}<\gamma_{2}^{i+1}<\cdots<\gamma_{2l}^{i+1}=\gamma_{l}^{i}$,
and hence the left-hand part of \eqref{eq:set_size} is proved. On
the other hand, $\gamma_{l}^{i}=\gamma_{2l}^{i+1}<\gamma_{2l+2}^{i+1}<\cdots<\gamma_{2\mathcal{L}-2}^{i+1}$,
and \eqref{eq:pmu_approx} implies that $\gamma_{2l+1}^{i+1}\geq\gamma_{2l}^{i+1}$.
As a result, $\gamma_{k}^{i+1}\geq\gamma_{l}^{i}$ for $k\geq2l$,
and the right-hand part of \eqref{eq:set_size} is proved.\end{IEEEproof}

\subsection{Double Thresholding Scheme}

Based on the path metric properties presented in Proposition 1, the
DTS is proposed for a fast LPO. It finds the $\mathcal{L}$ approximately
smallest path metrics from the $2\mathcal{L}$ inputs to form the
surviving path metric set $\Psi$.

\textit{Double Thresholding Scheme:} Assuming the $\mathcal{L}$ path
metrics $\gamma_{l}^{i}$ ($l=0,1,\ldots,\mathcal{L}-1$) input to
the PMU satisfy \eqref{eq:sort_assumption}, two threshold values,
one the \textit{acceptance threshold} ($AT$) and the other the \textit{rejection
threshold} ($RT$), can be determined, and they are given as
\begin{equation}
\left[AT,RT\right]=\left[\gamma_{\mathcal{L}/2}^{i},\gamma_{\mathcal{L}-1}^{i}\right].\label{eq:DT}
\end{equation}
The LPO for $\gamma_{k}^{i+1}$ ($k=0,1,\ldots,2\mathcal{L}-1$) is
then summarized as follows: \begin{enumerate} [leftmargin=10ex]%[leftmargin=*,topsep=0pt,itemsep=-1ex,partopsep=1ex,parsep=1ex]
\item [DTS.1)] if $\gamma_{k}^{i+1}<AT$, $\gamma_{k}^{i+1}\in\Psi$;
\item [DTS.2)] if $\gamma_{k}^{i+1}>RT$, $\gamma_{k}^{i+1}\notin\Psi$; and
\item [DTS.3)] if $AT\leq \gamma_{k}^{i+1}\leq RT$, it is randomly chosen to be included in $\Psi$ such that $\left|\Psi\right|=\mathcal{L}$.
\end{enumerate}  Finally, the path extensions with the path metrics $\gamma_{k}^{i+1}$s
that are inside $\Psi$ are kept and the rest of the path extensions
are pruned. \begin{figure} 
\centering 
\subfloat[]{\includegraphics{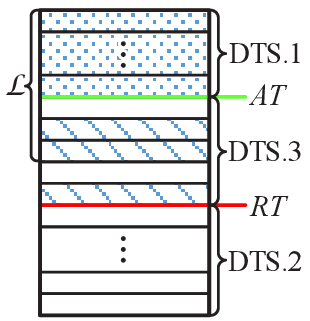}} \!
\subfloat[]{\includegraphics{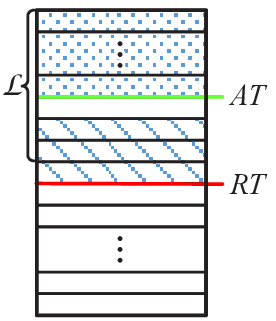}} \!
\subfloat[]{\includegraphics{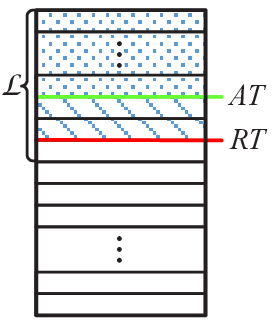}}
\caption{Double thresholding scheme. (a) $RT=\gamma_{\mathcal{L}-1}^{i}$. (b) A smaller $RT$. (c) A too small $RT$.} 
\end{figure}

The operation of the DTS is illustrated in Fig. 5. Assuming the $2\mathcal{L}$
path metrics $\gamma_{k}^{i+1}$ ($k=0,1,\ldots,2\mathcal{L}-1$)
are sorted in ascending order, the top $\mathcal{L}$ path metrics
are the smallest. Hence, they are the elements of $\Psi$ if an exact
sorting method is used for the LPO. On the other hand, when the DTS
is used, the shaded path metrics are the elements of $\Psi$. 

From Proposition 1, DTS.1 ensures that at least $\mathcal{L}/2$ path
metrics are picked and they are the smallest among all $2\mathcal{L}$
path metrics. So these path metrics are in the original exactly-sorted
$\Psi$. Therefore, based on DTS.1, the performance of the resulting
LSCD with list size $\mathcal{L}$ would not be worse than that of
the LSCD with a list size $\mathcal{L}/2$ based on the exact sorting
method.

From Proposition 1, $\left|\Omega\left(RT\right)\right|\geq\mathcal{L}-1$,
and \eqref{eq:pmu_approx} implies $\gamma_{2\mathcal{L}-2}^{i+1}=RT$.
Hence, at least $\mathcal{L}$ $\gamma_{k}^{i+1}$s are less than
or equal to $RT$. It also means that at most $\mathcal{L}$ path
metrics are greater than $RT$. Therefore, DTS.2 efficiently excludes
at most the $\mathcal{L}$ largest path metrics and these are surely
not in the original exactly-sorted $\Psi$. Finally, as shown in Fig.
5(a), when the number of path metrics picked by DTS.1 is smaller than
$\mathcal{L}$, DTS.3 randomly chooses the metrics from the remaining
$\gamma_{k}^{i+1}$s to fill up the decoding list such that $\left|\Psi\right|=\mathcal{L}$.
\begin{algorithm2e}[t] 
\caption{Procedure of the low-latency LSCD}
$L\leftarrow 1$\tcp*[r]{initialize the actual size of the candidate list}
\For {$i=0,1,\cdots N-1$}{
\For (\tcp*[f]{given the $l^{\textrm{th}}$ decoding path $\hat{\mathbf{u}}_{0}^{i-1}$}) {$l=0,1,\cdots L-1$}{
update $\Lambda_{l}^{i}$ with \eqref{eq:f_function}-\eqref{eq:partial_sum}\tcp*[r]{SCD operation}
\uIf {$i\in\mathcal{A}^{c}$} {
extend $\hat{\mathbf{u}}_{0}^{i-1}$ to $\hat{\mathbf{u}}_{0}^{i}$ with $\hat{u}_{i}\leftarrow 0$\;
update $\gamma_{l}^{i+1}$ from $\gamma_{l}^{i}$ and $\Lambda_{l}^{i}$ with $\hat{u}_{i}\leftarrow 0$\tcp*[r]{PMU in \eqref{eq:pm_update}}
}
\uElseIf{$i\in\mathcal{A}_{r}$}{
extend $\hat{\mathbf{u}}_{0}^{i-1}$ to $\hat{\mathbf{u}}_{0}^{i}$ with $\hat{u}_{i}\leftarrow \Theta\left(\Lambda_{l}^{i}\right)$\;
update $\gamma_{l}^{i+1}$ from $\gamma_{l}^{i}$ and $\Lambda_{l}^{i}$ with $\hat{u}_{i}\!\leftarrow\! \Theta\!\left(\!\Lambda_{l}^{i}\right)$\tcp*[r]{PMU in \eqref{eq:pm_update}}
}
\Else (\tcp*[f]{$i\in\mathcal{A}_{u}$}) {
extend $\hat{\mathbf{u}}_{0}^{i-1}$ to two path $\hat{\mathbf{u}}_{0}^{i}$s with $\hat{u}_{i}\leftarrow 0/1$\;
update $\gamma_{l}^{i+1}$s from $\gamma_{l}^{i}$ and $\Lambda_{l}^{i}$ with $\hat{u}_{i}\!\in\!\left\{ 0,1\right\}$\tcp*[r]{PMU in \eqref{eq:pm_update}}
$L\leftarrow L+1$;
}
}
\If (\tcp*[f]{DTS for LPO}) {$L>\mathcal{L}$}{
update $\Psi$ from $\gamma_{l}^{i+1}$s, $AT$, and $RT$, and reserve corresponding $\hat{\mathbf{u}}_{0}^{i}$s\;
$L\leftarrow \mathcal{L}$;
}
find $AT$ and $RT$ from the updated $\Psi$;
}
\Return {$\hat{\mathbf{u}}_{N}$} passed the CRC check;
\end{algorithm2e}

Compared with the exact-sorting method, the performance of the DTS
is potentially degraded due to DTS.3. As shown in Fig. 5(a), some
of larger of the $\mathcal{L}$ smallest path metrics may not be chosen
by DTS.3, and this happens when the number of path metrics accepted
by DTS.1 and that excluded by DTS.2 are both fewer than $\mathcal{L}$.
Therefore, to improve the performance of the DTS, a larger $AT$ or
a smaller $RT$ can be used. If the $AT$ is increased, it is possible
that more than $\mathcal{L}$ path metrics are accepted by DTS.1.
Also, as will be discussed in the next section, in order to reduce
the number of comparisons, our proposed architecture does not explicitly
generate the $AT$ value for comparison. Hence, in this work, a smaller
$RT$, e.g., $RT=\gamma_{l}^{i}$ ($l<\mathcal{L}-1$), is used to
improve the performance. As indicated in Fig. 5(b), a smaller $RT$
excludes more path metrics, and hence the path metric chosen by DTS.3
is more likely to be one of the $\mathcal{L}$ smallest metrics. On
the other hand, with a smaller $RT$, it is possible that more than
$\mathcal{L}$ path metrics will be excluded by DTS.2. As shown in
Fig. 5(c), this results in a list size smaller than $\mathcal{L}$.
Hence, if the $RT$ is reduced by too much, the performance of the
LSCD will also be degraded. In the next section, we propose an architecture
that can use a smaller $RT$ value while guaranteeing to generate
a list with size $\mathcal{L}$.

The overall procedure of the proposed low-latency LSCD based on SE
and DTS is summarized in Algorithm 2. Lines 8-14 execute the SE method
discussed in Section III and Lines 15-18 describe the DTS. From the
hardware implementation perspective, since now we only need to compare
the $2\mathcal{L}$ input path metric values with fixed threshold
values, the DTS can be executed in parallel, without a large increase
in computation complexity. Therefore, the logic delay is much smaller
than that of the exact sorting method and the overall latency of the
LPO is reduced. In the next section, a VLSI architecture implementing
Algorithm 2 will be discussed in detail. 
\begin{figure}
\centering{}\includegraphics{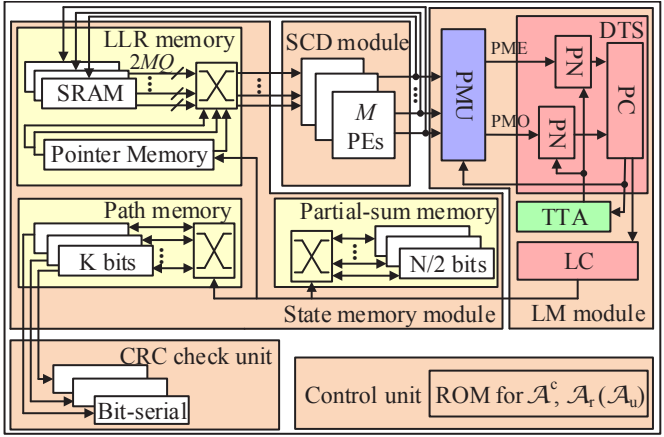}\caption{Top-level architecture of the low-latency LSCD.}
\end{figure}

\section{Low-Latency LSCD Architecture}

The top-level architecture of the proposed LSCD is shown in Fig. 6.
It mainly consists of five modules: the \textit{SCD module}, the \textit{state
memory module}, the \textit{LM module}, the \textit{CRC check unit},
and the \textit{control unit}. The SCD module is composed of $\mathcal{L}$
independent semi-parallel SCDs, each using $M$ ($M<N/2$) processing
elements (PEs) for the $f$ and $g$ function evaluation \cite{semi_parallel_Gross},
\cite{psn}. The CRC check unit contains $\mathcal{L}$ bit-serial
units computing the CRC check of each decoding path. As shown in \cite{sTSP_EPFL},
the latency of the CRC check unit is masked by that of the LSCD and
hence can be neglected. A $2N$ bit ROM is used to store the flags
to indicate whether $u_{i}$ is a frozen bit, a reliable information
bit, or an unreliable information bit, and this is used by the control
unit to generate the corresponding control signals to each block.
In the rest of this section, the state memory module and the LM module
are discussed in detail.

\subsection{State Memory Module}

Similar to the architecture in \cite{TCASII_EPFL}, the state memory
module is composed of three memories: the \textit{LLR memory}, storing
the intermediate $L_{j}^{t}$s ($0\leq j<2^{t}$, $0\leq t\leq n$)
of each SCD; the \textit{partial-sum memory}, storing the partial-sums
of each SCD \cite{psn}; and the \textit{path memory}, storing the
$\mathcal{L}$ decoding paths.

As discussed in \cite{semi_parallel_Gross} and \cite{psn}, a semi-parallel
SCD with $M=2^{m}$ processing elements uses a dual-port SRAM to store
the intermediate LLR operands at every decoding stage. It consists
of $2\left(\frac{N}{M}+m\right)$ words with $MQ$ bits each (i.e.,
an overall size of $2\left(N+mM\right)Q$ bits), where $Q$ is the
number of quantization bits for the LLR values. In every cycle, two
words are needed for the corresponding $f$ and $g$ node execution
and one word of the $M$ LLR values is generated and stored back.
$NQ$ bits of memory are used to store the channel input LLR $L_{i}^{n}$
($0\leq i<N$) and the remaining $\left(N+2mM\right)Q$ bits are used
for the intermediate output LLR $L_{j}^{t}$ ($0\leq j<2^{t}$, $0\leq t<n$).
To support the operation of $\mathcal{L}$ parallel SCDs, $\mathcal{L}$
SRAMs are needed for the LLR memory. Since the channel input $L_{i}^{n}$s
are the same for all $\mathcal{L}$ SCDs, they can be stored in the
first SRAM, while the size of the other SRAMs is reduced to $\left(N+2mM\right)Q$
bits each. As a result, the overall size of the LLR memory is $\left[\left(\mathcal{L}+1\right)N+2\mathcal{L}mM\right]Q$
bits.%
\footnote{As discussed in \cite{semi_parallel_Gross}, for an easy memory layout
and a simple connection between the memory and the PEs, every word
of the memory has the same bit width. For each SCD, the memory location
for storing $L_{j}^{t}$s at stage $t$, where $0\leq t\leq m$, has
$\left(2mM+1\right)Q$ unused bits and hence for an LSCD with list
size $\mathcal{L}$, there is an overall unused overhead of $\left(2mM+1\right)\mathcal{L}Q$
bits.%
}

As shown in \cite{psn}, $N/2$ bits of partial-sums are stored for
the $g$ function evaluation for one SCD. Hence, the size of the partial-sum
memory in the LSCD is $\mathcal{L}N/2$ bits. The size of the path
memory is $\mathcal{L}K$ bits, as each of the $\mathcal{L}$ decoding
paths has $K$ information bits (the values of the $N-K$ frozen bits
are pre-known and need not be stored). Since the sizes of the partial-sum
memory and the path memory are much smaller than that of the LLR memory,
they are implemented using registers and organized into $\mathcal{L}$
register blocks with equal size, as shown in Fig. 6.

For LSCD, each SCD expands a decoding path into two when an information
bit is decoded. The two paths can both be kept in or excluded from
the surviving candidate list. That means an SCD used for the decoding
of a path stored in a certain SRAM in this decoding cycle may be assigned
to decode another path stored in another SRAM in the next decoding
cycle. Therefore, we need to re-align the connection between the state
memory and the SCD in each decoding cycle. As shown in Fig. 6, for
the partial-sum memory and the path memory, $\mathcal{L}\times\mathcal{L}$
crossbars are used for moving the data for the alignment. For the
LLR memory, since the size is very large and moving the contents has
a large timing and power overhead, the lazy copy method, which uses
a pointer to manipulate the alignment instead of physically moving
the data content, is introduced in \cite{list} and \cite{TCASII_EPFL}.
As shown in Fig. 6, an $\mathcal{L}\times\mathcal{L}$ crossbar with
port width $2MQ$ bits is used to direct the memory contents to the
corresponding SCD hardware. The control signals of this crossbar are
generated by the \textit{pointer memory} updated by the LM module,
and the details of the updating logic have been presented in \cite{TCASII_EPFL}
and \cite{sTSP_EPFL}. The size of the pointer memory is $\mathcal{L}\times\left(n-1\right)\times\log_{2}\mathcal{L}$
bits, and the memory is implemented with registers. 
\begin{figure}
\begin{centering}
\includegraphics{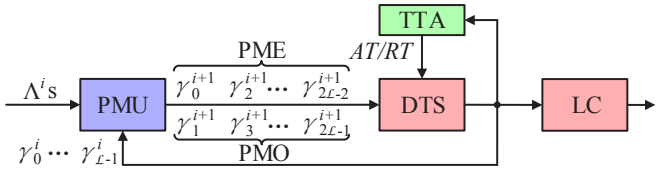}
\par\end{centering}

\caption{The data path of the LM module using the DTS.}
\end{figure}

\subsection{List Management Module}

The LM module implements the LM operation shown in Fig. 3. Fig. 7
shows the data path when the DTS is used for the LPO. It mainly consists
of four components: the \textit{threshold-tracking architecture} (TTA),
the PMU block, the DTS block, and the lazy copy (LC) block. Specifically,
the PMU block executes the PMU operation in Fig. 3, and the DTS block
together with the LC block implements the LPO shown in Fig. 3. The
TTA calculates the thresholds to support the operation of the DTS
block. As shown in Fig. 7, after decoding $u_{i-1}$, the path metrics
of the $\mathcal{L}$ surviving decoding paths are $\gamma_{l}^{i}$
($l=0,1,\ldots,\mathcal{L}-1$). In decoding $u_{i}$ ($i\in\mathcal{A}_{u}$),
the $\mathcal{L}$ SCDs generate $\mathcal{L}$ output LLRs $\Lambda_{l}^{i}$s,
and the PMU block generates the path metrics of the $2\mathcal{L}$
extended paths. After this, the DTS block finds the $\mathcal{L}$
almost-the-best path metrics $\gamma_{l}^{i+1}$ ($l=0,1,\ldots,\mathcal{L}-1$)
and their corresponding decoding paths. Based on the information on
path removal and survival, the LC block manipulates the memory contents
in the state memory module, and its logic has been discussed in \cite{TCASII_EPFL}
and \cite{sTSP_EPFL}. Running in parallel with the LC block, the
TTA block calculates the values of $AT$ and $RT$ from the surviving
$\gamma_{l}^{i+1}$s, and they will be used by the DTS block for the
decoding of the next bit. In the following, the architectures for
the PMU, TTA, and DTS blocks are presented in detail.

\subsubsection{PMU Block in the List Management Module}

The PMU block expands and updates the path metrics based on \eqref{eq:pmu_approx}.
Its $2\mathcal{L}$ outputs $\gamma_{l}^{i+1}$ ($l=0,1,\ldots,2\mathcal{L}-1$)
are divided into two groups: path metrics with an even index (PME),
i.e. $\gamma_{j}^{i+1}$ ($j=0,2,\ldots,2\mathcal{L}-2$), and path
metrics with an odd index (PMO), i.e. $\gamma_{k}^{i+1}$ ($k=1,3,\ldots,2\mathcal{L}-1$).
From \eqref{eq:pmu_approx}, no extra hardware is required to generate
the path metrics in the PME as $\gamma_{j}^{i+1}=\gamma_{j/2}^{i}$
when $j$ is an even number. On the other hand, $\mathcal{L}$ adders
are needed in the PMU block to generate the path metrics in the PMO
as $\gamma_{k}^{i+1}=\gamma_{\left(k-1\right)/2}^{i}+\left|\Lambda_{\left(k-1\right)/2}^{i}\right|$
when $k$ is an odd number.

\subsubsection{TTA in the List Management Module}

The TTA is responsible for calculating the acceptance threshold $AT$
and the rejection threshold $RT$ for the DTS to work. The $AT$ and
$RT$ values for decoding bit $u_{i}$ are generated from $\gamma_{l}^{i}$
($l=0,1,\ldots,\mathcal{L}-1$), which are the $\mathcal{L}$ surviving
path metrics at bit $u_{i-1}$, as shown in Fig. 7. The architecture
of the TTA is shown in Fig. 8. In addition to the generation of $AT$
and $RT$, as shown in Fig. 8, the TTA also outputs the partially-sorted
$\gamma_{l}^{i}$s. The smallest $\mathcal{L}/2$ path metrics are
on the top and the largest $\mathcal{L}/2$ path metrics, which are
exactly-sorted, are at the bottom. The details of the TTA operations
are as follows. 
\begin{figure}
\begin{centering}
\includegraphics{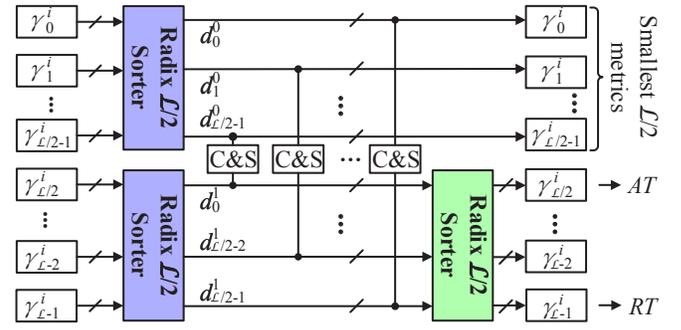}
\par\end{centering}

\caption{Threshold-tracking architecture.}
\end{figure}

The $\mathcal{L}$ input path metrics are evenly divided into two
groups. Each group is then sorted by a radix-$\mathcal{L}/2$ sorter
\cite{Parallel_Sorting}. Therefore, their outputs $d_{j}^{k}$ ($j=0,1,\ldots,\mathcal{L}/2-1$;
$k=0,1$) satisfy $d_{0}^{k}\leq d_{1}^{k}\leq\cdots\leq d_{\mathcal{L}/2-1}^{k}$,
for $k=0$ and 1. Similar to \cite{ISCAS_Lehigh}, $\mathcal{L}/2$
comparing-and-swapping (C\&S) elements take pairs of the output values
of the sorters $\left(d_{j}^{0},d_{\mathcal{L}/2-1-j}^{1}\right)$,
$j=0,1,\ldots,\mathcal{L}/2-1$, as their inputs and direct the smaller
value to the upper output and the larger value to the lower output.
As a result, the outputs of the C\&S array are partially sorted, where
the top $\mathcal{L}/2$ outputs are guaranteed to be smaller than
or equal to the lower $\mathcal{L}/2$ outputs. For an easier implementation
of the DTS architecture, the lower $\mathcal{L}/2$ outputs are further
exactly sorted by another radix-$\mathcal{L}/2$ sorter. The reason
for this will be discussed in the next sub-section. From the discussion
in Section IV-B, the first element in the lower $\mathcal{L}/2$ sorted
output path metric $\gamma_{\mathcal{L}/2}^{i}$ in Fig. 8 is $AT$.
In fact, we do not need to know the value of $AT$. The group of path
metrics that satisfies the $AT$ check can be directly obtained from
the top $\mathcal{L}/2$ outputs of the TTA. This will be discussed
in more detail in the next sub-section. Moreover, $RT$ can be chosen
from the lower $\mathcal{L}/2$ sorted output path metrics of the
TTA. For example, the $RT$ used in \eqref{eq:DT} is the last output
path metric of the TTA. The TTA requires an exact sorting of $\mathcal{L}/2$
elements. For other LSCD architectures that use exact sorting for
list pruning, the input size of the sorter is $2\mathcal{L}$ instead
of $\mathcal{L}/2$. So the complexity of the proposed TTA is much
smaller. In addition, the TTA is executed in parallel with the execution
of $f$ or $g$ nodes and the PMU for the decoding of the next bit,
and hence the latency is hidden and no extra cycle is added to the
overall latency.

\subsubsection{DTS Block in the List Management Module}

As shown in Fig. 7, when we decode bit $u_{i}$, the DTS takes the
two groups of path metrics (PME and PMO) output from the PMU block
as input. The $AT$ and $RT$ values obtained from the TTA are used
as the threshold values for the DTS operation.

As shown in Fig. 6, the path metrics in the PME and PMO are firstly
passed to two \textit{permutation networks} (PNs), respectively. Since
each partially-sorted path metric output of the TTA corresponds to
the generation of one path metric element in the PME and another in
the PMO, according to \eqref{eq:pmu_approx}, the elements in the
PME and PMO are permutated based on the sorted-order of their parent
path metrics in the TTA output. For example, if the orders of the
outputs of the TTA are $\gamma_{3}^{i}$, $\gamma_{2}^{i}$, $\gamma_{0}^{i}$,
and $\gamma_{1}^{i}$ for $\mathcal{L}=4$, the orders of the PME
and PMO after permutation are $\left[\gamma_{6}^{i+1},\gamma_{4}^{i+1},\gamma_{0}^{i+1},\gamma_{2}^{i+1}\right]$
and $\left[\gamma_{7}^{i+1},\gamma_{5}^{i+1},\gamma_{1}^{i+1},\gamma_{3}^{i+1}\right]$,
respectively. Since the first $\mathcal{L}/2$ outputs of the TTA
are smaller than $AT$ and $\gamma_{2l}^{i+1}=\gamma_{l}^{i}$, the
first $\mathcal{L}/2$ elements in the permutated PME are all smaller
than $AT$. Similarly, as $AT$ is the smallest value among the last
$\mathcal{L}/2$ outputs of the TTA, the last $\mathcal{L}/2$ elements
in the permutated PME are all greater than or equal to $AT$. 
\begin{figure}
\begin{centering}
\includegraphics{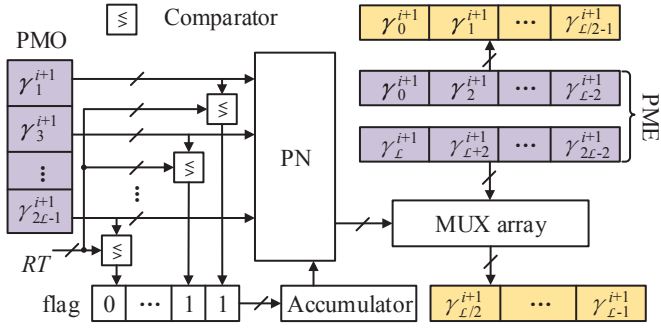}
\par\end{centering}

\caption{Architecture of the pruning and copying (PC) block.}
\end{figure}

After permutation, the elements of the PME and PMO are passed to the
\textit{pruning and copying} (PC) block to determine the $\mathcal{L}$
surviving paths. The architecture of the PC is shown in Fig. 9. From
the above discussion, the first $\mathcal{L}/2$ elements in the PME
are definitely smaller than $AT$ and hence will be included in the
surviving set $\Psi$. To fill up the remaining $\mathcal{L}/2$ elements
in $\Psi$, as discussed in Section IV, we need to compare the last
$\mathcal{L}/2$ elements in the PME and the elements in the PMO with
$AT$ and $RT$. Random inclusion or exclusion has to be done if the
number of elements passing the two threshold checks is not exactly
equal to $\mathcal{L}/2$. To reduce the number of comparisons and
also avoid the random inclusion/exclusion, which will complicate the
hardware implementation, we propose a different method to select the
remaining $\mathcal{L}/2$ elements in $\Psi$. We temporarily accept
the last $\mathcal{L}/2$ elements in the PME first. We then compare
the elements in the PMO with a fixed $RT$ value using $\mathcal{L}$
comparators. A flag equal to 1 is generated if the corresponding path
metric is not greater than $RT$. Note that this $RT$ value is smaller
than that stated in \eqref{eq:DT} in order to prune out more paths
with larger metric values. All the flags are then added up by an accumulator
to decide how many path metrics are not greater than $RT$. Carry-save
adders and adder tree are used to reduce the delay of the accumulator.
Let $k$ be the output of the accumulator. Then the largest $k$ elements
of the last $\mathcal{L}/2$ elements in the PME are replaced by the
$k$ path metrics in the PMO that are not greater than $RT$. Note
that since the last $\mathcal{L}/2$ elements in the PME are exact-sorted
in order, we simply pick the last $k$ elements in the set for replacement.
If $k$ is larger than $\mathcal{L}/2$, we just take the first $\mathcal{L}/2$
elements in the PMO that pass the $RT$ test to replace the last $\mathcal{L}/2$
elements in the PME in $\Psi$. 

The DTS architecture presented in Fig. 9 has two advantages over the
DTS operation discussed in Section IV-B. Firstly, a much smaller $RT$
can be used to exclude more paths with large metric values. Even when
a smaller $RT$ is used, we can still guarantee at any time that the
candidate list of the LSCD has $\mathcal{L}$ decoding paths. In the
worst case, when all the path metrics in the PMO are greater than
$RT$, we will keep the last $\mathcal{L}/2$ elements in the PME
in the surviving path list. Secondly, since the last $\mathcal{L}/2$
elements in the PME are already sorted by the TTA, we always replace
the worst elements in the PME. This is better than randomly selecting
a path to replace as the probability of the last few elements of the
PME in the actual surviving path set is low. As a result, the error-correcting
performance of the DTS is improved by using the architecture shown
in Fig. 9, and we denote this as DTS-Advance.
\begin{figure}
\begin{centering}
\subfloat[]{\begin{centering}
%\footnotesize
\small
\setlength\tabcolsep{10pt}\setlength{\extrarowheight}{1pt}
\begin{tabular}{c"c|c|c|c} 
\hlinewd{1pt}  
cycle & 0 & 1 & 2 & 3\tabularnewline 
\hlinewd{1pt}  
SCD & \cellcolor{black!25}$f_{0}^{1}$ & \cellcolor{black!25}$f_{0}^{0}$ & \cellcolor{black!25}$g_{0}^{0}$ & \cellcolor{black!25}$g_{0}^{1}$\tabularnewline 
\hlinewd{1pt}  
\end{tabular}
\par\end{centering}

}
\par\end{centering}

\begin{centering}
\subfloat[]{\begin{centering}
%\footnotesize
\small
\setlength\tabcolsep{5pt}\setlength{\extrarowheight}{1pt}
\begin{tabular}{c"c|c|c|c|c|c|c|c>{\centering}p{1.1cm}} 
\hlinewd{1pt}  
cycle & \multicolumn{2}{>{\centering}p{1.1cm}|}{0} & \multicolumn{2}{>{\centering}p{1.1cm}|}{1} & \multicolumn{2}{>{\centering}p{1.1cm}|}{2} & \multicolumn{2}{>{\centering}p{1.1cm}|}{3} & 4\tabularnewline 
\hlinewd{1pt}  
SCD & \cellcolor{black!25}$f_{0}^{0}$ &  & \multicolumn{2}{c|}{} & \cellcolor{black!25}$g_{0}^{0}$ &  & \multicolumn{2}{c|}{} & \cellcolor{black!25}$g_{0}^{1}$\tabularnewline 
\hline 
PMU &  & \cellcolor{blue!25}$\gamma_{l}^{1}$ & \multicolumn{2}{c|}{} &  & \cellcolor{blue!25}$\gamma_{l}^{2}$ & \multicolumn{2}{c|}{} & \tabularnewline 
\hline 
DTS &  \multicolumn{2}{c|}{}  & \cellcolor{red!25}$\gamma_{l}^{1}$ &  & \multicolumn{2}{c|}{} & \cellcolor{red!25}$\gamma_{l}^{2}$ & \multicolumn{1}{c|}{} & \tabularnewline 
\hline  
LC &  \multicolumn{2}{c|}{}  &  & \cellcolor{red!25}LC & \multicolumn{2}{c|}{} &  & \multicolumn{1}{c|}{\cellcolor{red!25}LC} & \tabularnewline 
\hline  
TTA &  \multicolumn{2}{c|}{}  &    & \multicolumn{3}{c|}{\cellcolor{green!25}TTA} &    & \multicolumn{2}{c}{\cellcolor{green!25}TTA}\tabularnewline 
\hlinewd{1pt}  
\end{tabular}
\par\end{centering}

}
\par\end{centering}

\caption{Timing diagram example of the low-latency LSCD architecture. (a) Timing
diagram of decoding $u_{0}$ and $u_{1}$ for $N=4$ polar codes.
(b) Timing diagram of decoding $u_{0}$ and $u_{1}$ with the LSCD.}
\end{figure}

\subsection{Decoding Latency of the Proposed LSCD Architecture}

Fig. 10(a) shows the timing diagram of decoding $u_{0}$ and $u_{1}$
in the scheduling tree of Fig. 4 using a single SCD. When LSCD is
used, additional cycles are required for the path metric updating
and list pruning. Fig. 10(b) shows the timing diagram of decoding
$u_{0}$ and $u_{1}$ with the proposed LSCD architecture,%
\footnote{For illustration, it is assumed that the list is already full of $\mathcal{L}$
decoding paths in the beginning.%
} where the detailed timing of the list management (LM) component is
also shown. Specifically, $\gamma_{l}^{i}$ in the PMU and DTS denotes
the generation of $2\mathcal{L}$ path metrics output from $\mathcal{L}$
input path metrics in the PMU block and finding the $\mathcal{L}$
surviving path metrics from the $2\mathcal{L}$ path metric candidates
in the DTS block, respectively. Compared with the architecture presented
in \cite{icassp}, the processing element data path is optimized and
the PMU block is executed in the same clock cycle with the leaf $f/g$
node execution of the SCD operation. Moreover, the LPO implemented
by the DTS and the lazy copying (LC) blocks are done in the same clock
cycle. Due to the data dependency, the TTA operation for finding the
threshold values for the next bit is executed when the DTS for the
current bit is finished and it is hidden in the cycle where the leaf
$f/g$ nodes are executed. As a result, by using the DTS for the LPO,
only one additional cycle is introduced for each LM operation.

From \cite{semi_parallel_Gross}, the decoding latency (i.e., the
time to traverse the scheduling tree) of a semi-parallel SCD using
$M$ PEs is equal to $2N+\frac{N}{M}\log_{2}\left(\frac{N}{4M}\right)$
clock cycles. Hence, the overall latency of the LSCD architecture
is
\begin{equation}
D=3N+\frac{N}{M}\log_{2}\left(\frac{N}{4M}\right).\label{eq:general_latency}
\end{equation}

As discussed in Section III, when the SE method is used, if $u_{i}$
is a reliable bit, i.e. $i\in\mathcal{A}_{r}$, the operation of the
PMU and the LPO after the decoding of bit $u_{i}$ are not required.
Moreover, the LPO for the frozen bit is not executed either. Hence,
the latency in \eqref{eq:general_latency} can be reduced. The latency
is further reduced by considering two source bits at a time. A source-bit
couple is defined as $\left(u_{2i},u_{2i+1}\right)$, with $i\in\left\{ 0,1,\ldots,N/2-1\right\} $.
\begin{table}
\caption{Different Latency Reduction Cases for Source-Bit Couple}

\centering{}%\footnotesize
\small
\setlength\tabcolsep{8pt}\setlength{\extrarowheight}{1pt}
\begin{tabular}{c"c|c|c} 
\hlinewd{1pt}
\backslashbox{$a_r$~}{$a_f$~~} & 2 & 1\footnote{From the properties of polar codes, if $a_{f}=1$ in $\left(u_{2i},u_{2i+1}\right)$, then $2i\in\mathcal{A}^{c}$ and $2i+1\in\mathcal{A}$. }  & 0\tabularnewline 
\hlinewd{1pt} 
2 & n/a & n/a & Case I\tabularnewline 
\hline  
1 & n/a & Case II & Case III\tabularnewline 
\hline  0 & Case IV & Case V & Case VI\tabularnewline 
\hlinewd{1pt}  
\end{tabular}
\end{table}
Based on the types of bits of $u_{2i}$ and $u_{2i+1}$, the source-bit
couples can be categorized into six cases, which are summarized in
Table I, where $a_{f}$ and $a_{r}$ denote the number of frozen bits
and reliable bits in a source-bit couple, respectively. Without loss
of generality, we use the couple $\left(u_{0},u_{1}\right)$ and its
decoding timing diagram in Fig. 10 for illustration in the following
discussion.

\subsubsection{Case I}

Both $u_{0}$ and $u_{1}$ are reliable information bits. Hence, the
LM operation after decoding each bit is saved. Moreover, since the
PMU operation is not needed, the output LLRs $\Lambda^{0}$ and $\Lambda^{1}$
are not needed, and hence the leaf nodes of the scheduling tree, $f_{0}^{0}$
and $g_{0}^{0}$, are not executed. For the $l^{\textrm{th}}$ decoding
path, the values of $\left(\hat{u}_{0},\hat{u}_{1}\right)$ on its
path extension are determined by the hard decision of the SCD assigned
for that decoding path and they are given by $\left[\Theta\left(L_{0}^{1}\right),\Theta\left(L_{1}^{1}\right)\right]\mathbf{F}$,
where $L_{0}^{1}$ and $L_{1}^{1}$ are the LLRs from the parent node
of the SCD, i.e., $f_{0}^{1}$ in Fig. 4. 

Based on the above discussion, the operations in cycles 0 to 3 of
Fig. 10(b) are saved for Case I. Moreover, as part of the LM operation,
the TTA in cycle 4 is also not needed. As a result, four clock cycles
are saved for the Case I source-bit couple.

\subsubsection{Case II}

Bit $u_{0}$ is a frozen bit and $u_{1}$ is a reliable information
bit. The LPOs for both bits and the PMU operation for bit $u_{1}$
are not executed. However, the PMU for the frozen bit $u_{0}$ still
has to be executed, and it can be combined with the SCD operation
as follows: \setlength{\arraycolsep}{0.1em} 
\begin{equation}
\gamma_{l}^{2}=\begin{cases}
\begin{array}{l}
\gamma_{l}^{0}\\
\gamma_{l}^{0}+\min\left(\left|L_{0}^{1}\right|,\left|L_{1}^{1}\right|\right)
\end{array} & \begin{array}{l}
\textrm{if}\;\Theta\left(L_{0}^{1}\right)=\Theta\left(L_{1}^{1}\right),\\
\textrm{if}\;\Theta\left(L_{0}^{1}\right)\neq\Theta\left(L_{1}^{1}\right),
\end{array}\end{cases}\label{eq:frz_inf_se_pmu}
\end{equation}
\setlength{\arraycolsep}{5pt}where $l=0,1,\ldots,\mathcal{L}-1$.
Similar to Case I, $L_{0}^{1}$ and $L_{1}^{1}$ are the LLRs output
from node $f_{0}^{1}$. 

As a result, for the Case II source-bit couple, the leaf nodes of
the scheduling tree are not executed and the LM operations are simplified
to \eqref{eq:frz_inf_se_pmu}. The $l^{\textrm{th}}$ decoding path's
path extension $\left(\hat{u}_{0},\hat{u}_{1}\right)$ is given as
$\left(0,\Theta\left(L_{0}^{1}+L_{1}^{1}\right)\right)$. Specifically,
the PMU operation in \eqref{eq:frz_inf_se_pmu} is retimed and it
is executed in the same cycle with $f_{0}^{1}$. Thus, the corresponding
operations in cycles 0 to 3 are not needed. Different from that in
Case I, the TTA in cycle 4 has to be executed, as the path metrics
are changed by \eqref{eq:frz_inf_se_pmu}.

\subsubsection{Case III}

$u_{0}$ is an unreliable information bit and $u_{1}$ is a reliable
bit.%
\footnote{From the properties of polar codes, if $\left\{ 2i,2i+1\right\} \subset\mathcal{A}$
and $a_{r}=1$, then $2i\in\mathcal{A}_{u}$ and $2i+1\in\mathcal{A}_{r}$.%
} In this case, the operations of the PMU, the LPO, and the TTA after
decoding $u_{1}$ are not needed. Hence, one clock cycle (i.e., cycle
3 in Fig. 10(b)) is saved.

\subsubsection{Case IV}

Both $u_{0}$ and $u_{1}$ are frozen bits. The LPOs for both bits
are saved, and the PMU operations of the two bits are combined and
simplified as \cite{icassp}
\begin{equation}
\gamma_{l}^{2}=\gamma_{l}^{0}+\Theta\left(L_{0}^{1}\right)\cdot\left|L_{0}^{1}\right|+\Theta\left(L_{1}^{1}\right)\cdot\left|L_{1}^{1}\right|,\label{eq:frz_frz}
\end{equation}
where $l=0,1,\ldots,\mathcal{L}-1$, and $L_{0}^{1}$ and $L_{1}^{1}$
are the output LLRs of node $f_{0}^{1}$. Therefore, the leaf node
operations $f_{0}^{0}$ and $g_{0}^{0}$ of the SCD together with
the LM operations are simplified to \eqref{eq:frz_frz}. This PMU
operation is retimed and it is executed in the same cycle with $f_{0}^{1}$.
Hence, similar to Case II, four clock cycles are saved.

\subsubsection{Case V}

$u_{0}$ is a frozen bit and $u_{1}$ is an unreliable information
bit. This case is different from Case II, because the LM operation
is needed for $u_{1}$. Hence, only the LPO for $u_{0}$ can be eliminated
and one cycle is saved.

\subsubsection{Case VI}

Both $u_{0}$ and $u_{1}$ are unreliable information bits. Fig. 10(b)
depicts the timing of this case, and no latency reduction is achieved.
\begin{table}
\caption{Latency Reduction for Different Source-Bit Couple Cases}

\centering{}%\footnotesize
\small
\setlength\tabcolsep{7pt}\setlength{\extrarowheight}{1pt}
\begin{tabular}{c"c|c|c|c|c|c} 
\hlinewd{1pt}
Case & I & II & III & IV & V & VI\tabularnewline 
\hlinewd{1pt} 
Number of cycles reduced & 4 & 4 & 1 & 4 & 1 & 0\tabularnewline 
\hlinewd{1pt}  
\end{tabular}
\end{table}

Table II summarizes the latency reduction achieved by different source-bit
couple cases. As a result, the decoding latency of the proposed LSCD
architecture is given as
\begin{equation}
D_{\textrm{LSCD}}=D-4\left(\mathcal{N}_{\textrm{I}}+\mathcal{N}_{\textrm{II}}+\mathcal{N}_{\textrm{IV}}\right)-\left(\mathcal{N}_{\textrm{III}}+\mathcal{N}_{\textrm{V}}\right),\label{eq:lscd_latency}
\end{equation}
where $\mathcal{N}_{\alpha}$ denotes the number of source-bit couples
for Case $\alpha$ found in the polar codes. These values depend on
the frozen set $\mathcal{A}^{c}$ and the reliable set $\mathcal{A}_{r}$.
To achieve the timing specified in \eqref{eq:lscd_latency}, the PMU
block shown in Figs. 6 and 7 has to support the operation of \eqref{eq:frz_inf_se_pmu}
and \eqref{eq:frz_frz}, and it is easily achieved with additional
comparators and adders.

\section{Experimental Results}

In this section, to demonstrate the error-correcting performances
of the proposed SE method and DTS algorithm, an $\left(N,R,r\right)=\left(1024,1/2,16\right)$
polar code is simulated over a binary-input AWGN channel.%
\footnote{As stated in Section II-B, when 16-bit CRC code is used, the information
set $\mathcal{A}$ of polar codes is extended such that $K=\left|\mathcal{A}\right|=NR+r=528$.
When SCD is used to decode polar codes, CRC code is not used and hence
the size of $\mathcal{A}$ remains to be $K=512$ for a same code
rate of $R=1/2$. Specifically, the information set $\mathcal{A}$s
of both polar codes with $K=528$ and $K=512$ are optimized for $E_{b}/N_{0}=1.5$
dB.%
} Then, we present the implementation results of the proposed LSCD
architecture, and then compare them with those of other existing works.
\begin{figure}
\begin{centering}
\includegraphics{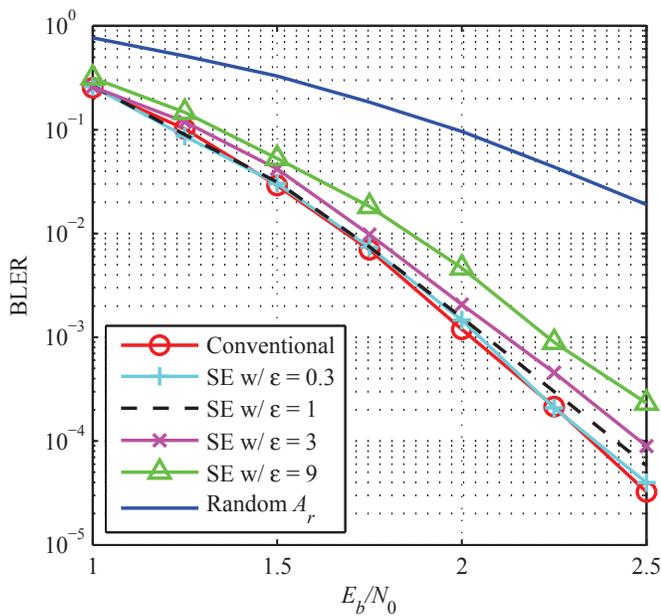}
\par\end{centering}

\caption{BLERs of LSCD using SE with different $\epsilon$s.}
\end{figure}

\subsection{Error-correcting Performance of the SE Method}

Fig. 11 shows the block-error rate (BLER) of different LSCD implementations
with a list size of $\mathcal{L}=16$. First the BLER of the conventional
LSCD, i.e., $P_{b}^{\textrm{LSCD}}$ in \eqref{eq:union_bound_SE}
and \eqref{eq:BLER_upper_bound}, is shown. The BLERs of the proposed
SE method with different sizes of the reliable set $\mathcal{A}_{r}$
are also shown. The size of $\mathcal{A}_{r}$ depends on the tolerable
performance degradation parameter $\epsilon$. In the simulation,
we use different $\epsilon$ values, ranging from 0.3 to 9 at $E_{b}/N_{0}=2.25$
dB. 
\begin{table}
\caption{The Cardinality of $\mathcal{A}_{r}$ and Decoding Latency for Different
$\epsilon$s}

\centering{}%\footnotesize
\small
\setlength\tabcolsep{8pt}\setlength{\extrarowheight}{1pt}
\begin{tabular}{c"c|c|c|c} 
\hlinewd{1pt}  
$\epsilon$@2.25 dB & 0.3 & 1 & 3 & 9\tabularnewline 
\hlinewd{1pt}  
$\left| \mathcal{A}_{r} \right|/\left| \mathcal{A} \right|$  & 72.35\% & 75.76\% & 78.98\% & 82.77\% \tabularnewline 
\hline
$D_{\textrm{LSCD}}$ (cycles) & 1462 & 1424 & 1381 & 1329 \tabularnewline 
\hlinewd{1pt}  
\end{tabular}
\end{table}

From Fig. 11, it can be seen that, for each given $\epsilon$, the
degradation in BLER of the LSCD using the SE method is close to the
upper bound predicted by \eqref{eq:union_bound_SE} and \eqref{eq:BLER_upper_bound}.
This indicates that the performance analysis in \eqref{eq:BLER_upper_bound}
well estimates the performance degradation introduced by the SE method
for a given reliable set $\mathcal{A}_{r}$. To investigate the relationship
between the latency reduction and the performance degradation of the
SE method, Table III summarizes the cardinality of $\mathcal{A}_{r}$
for different $\epsilon$s. Moreover, based on $\mathcal{A}^{c}$
and the corresponding $\mathcal{A}_{r}$, Table IV presents the number
of different source-bit couples for each $\epsilon$ value. Assuming
that the LSCD architecture proposed in Section V is used and each
SCD uses $M=64$ PEs, the last row of Table III compares the decoding
latency ($D_{\textrm{LSCD}}$) for different $\epsilon$s, based on
\eqref{eq:lscd_latency}. From Table III, we can see that for $\epsilon=0.3$,
more than 72\% of the information bits are included in set $\mathcal{A}_{r}$
and hence more than 72\% of the LPOs are saved by the corresponding
LSCD with SE. From Fig. 11, it is also shown that the performance
degradation introduced by the SE method with $\epsilon=0.3$ is negligible
compared with that of the conventional LSCD. If a larger $\epsilon$
is used, Table III shows that $\left|\mathcal{A}_{r}\right|$ is only
slightly increased, while the performance of the corresponding LSCD
is degraded significantly, as shown in Fig. 11. For example, when
$\epsilon=9$, the decoding latency is only reduced by 9\% compared
with that of $\epsilon=0.3$. Therefore, $\epsilon=0.3$ is used in
the SE method for our low-latency LSCD implementation. 

To verify the effectiveness of the method proposed in Section III
in finding set $\mathcal{A}_{r}$, we randomly choose 72.35\% information
bits in $\mathcal{A}$ to compose set $\mathcal{A}_{r}$. Fig. 11
shows its BLER using the SE method. It is shown that the performance
is greatly degraded from that using $\mathcal{A}_{r}$ generated from
our proposed method. 
\begin{figure}
\begin{centering}
\includegraphics{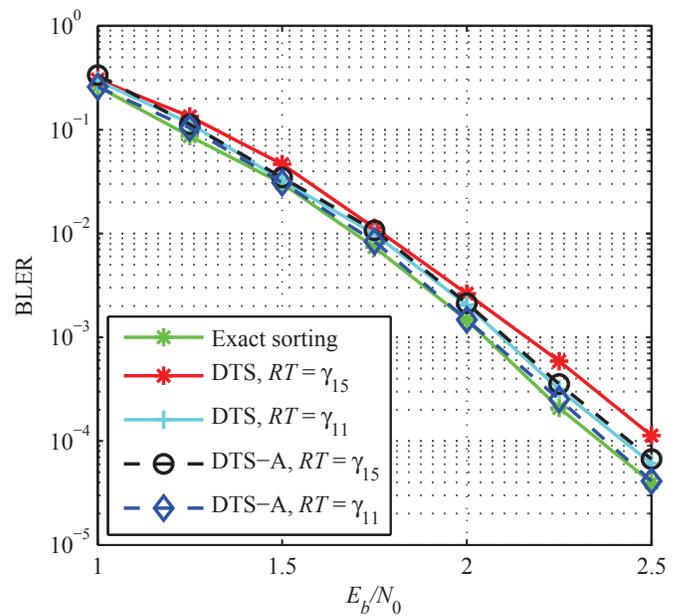}
\par\end{centering}

\caption{BLERs of LSCD using DTS and DTS-Advance with different $RT$ values.}
\end{figure}
 
\begin{table}
\caption{Source-Bit Couple Distribution for Different $\epsilon$s}

\centering{}%\footnotesize
\small
\setlength\tabcolsep{8pt}\setlength{\extrarowheight}{1pt}
\begin{tabular}{c"c|c|c|c|c|c} 
\hlinewd{1pt}
$\epsilon$ & $\mathcal{N}_{\textrm{I}}$ & $\mathcal{N}_{\textrm{II}}$ & $\mathcal{N}_{\textrm{III}}$ & $\mathcal{N}_{\textrm{IV}}$ & $\mathcal{N}_{\textrm{V}}$ & $\mathcal{N}_{\textrm{VI}}$\tabularnewline 
\hlinewd{1pt} 
0.3 & 158 & 0 & 66 & 224 & 48 & 16\tabularnewline 
\hline
1 & 168 & 0 & 64 & 224 & 48 & 8\tabularnewline 
\hline
3 & 176 & 5 & 60 & 224 & 43 & 4\tabularnewline 
\hline
9 & 186 & 11 & 54 & 224 & 37 & 0\tabularnewline 
\hlinewd{1pt}  
\end{tabular}
\end{table}
 
\begin{table*}
\caption{Synthesis Results Comparison of Different LSCD Architectures for $\left(N,R\right)=\left(1024,1/2\right)$
polar codes}

\centering{}%\footnotesize
\small
\setlength\tabcolsep{5pt}\setlength{\extrarowheight}{1pt}
\begin{tabular}{c"c|c|c|c|c|c|c} 
\hlinewd{1pt}  
 & This work & \cite{sTSP_EPFL} & \cite{ISCAS_Lehigh} & \cite{TVLSI_Parhi} & \cite{arxiv_Lehigh_Parallel} & \cite{ICASSP_EPFL} & \cite{TCASII_EPFL}\tabularnewline 
\hlinewd{1pt}  
PE number per SCD $M$ & \multicolumn{3}{c|}{64} & n/a & \multicolumn{3}{c}{64}\tabularnewline 
\hline
$K=\left|\mathcal{A}\right|$ & 528 & 528\footnote{A 16-bit CRC code is used with $\left(N,R\right)=\left(1024,1/2\right)$ polar codes in \cite{sTSP_EPFL}} & \multicolumn{5}{c}{512}\tabularnewline 
\hline
List size $\mathcal{L}$ & 16 & \multicolumn{2}{c|}{8} & \multicolumn{4}{c}{4}\tabularnewline 
\hlinewd{1pt} 
Technology & UMC 90 nm & TSMC 90 nm & 90 nm & ST 65 nm & TSMC 90 nm & UMC 90 nm & UMC 90 nm\tabularnewline 
\hline
Area ($\textrm{mm}^{2}$) & 7.47 & 3.85 & 8.64 & 2.14 & 1.669 & 1.743 & 3.53\tabularnewline 
\hline
Clock freq. (MHz) & 658 & 637 & 625 & 400 & 500 & 412 & 314\tabularnewline 
\hline
Throughput (Mbps) & 460 & 245 & 177 & 401 & 332 & 162 & 124\tabularnewline 
\hlinewd{1pt} 
\end{tabular}
\end{table*}

\subsection{Error-correcting Performance of the DTS}

Next the error-correcting performance of LSCD using the DTS to replace
exact sorting in the LPO is investigated. Simulations for the polar
code used in the previous sub-section are carried out. Fig. 12 shows
the BLERs of different LSCDs, including those using the DTS discussed
in Section IV and the DTS-Advance discussed in Section V. Comparisons
of the BLERs of the DTS using different $RT$ values are also shown.
Compared with the LSCD using the exact sorting method, when $\gamma_{\mathcal{L}-1}^{i}$
is used as $RT$, as stated in \eqref{eq:DT}, the LSCD using the
DTS introduces an SNR penalty of around 0.2 dB when the BLER is $10^{-4}$.
For the DTS-Advance discussed in Section V-B, the SNR loss is only
around 0.1 dB. Moreover, when a smaller $RT$ value is used, such
as $\gamma_{11}^{i}$ shown in Fig. 8, the performance degradation
of the DTS-Advance is negligible. However, when the same $RT$ is
used for the DTS, a performance loss of around 0.1 dB is recorded.
This is because fewer decoding paths are chosen by DTS.3 and the candidate
list is not full for most of the time. As a result, the DTS-Advance
with $RT=\gamma_{11}^{i}$ is used for a low-latency LPO in our LSCD
implementation. 
\begin{figure}
\centering{}\includegraphics{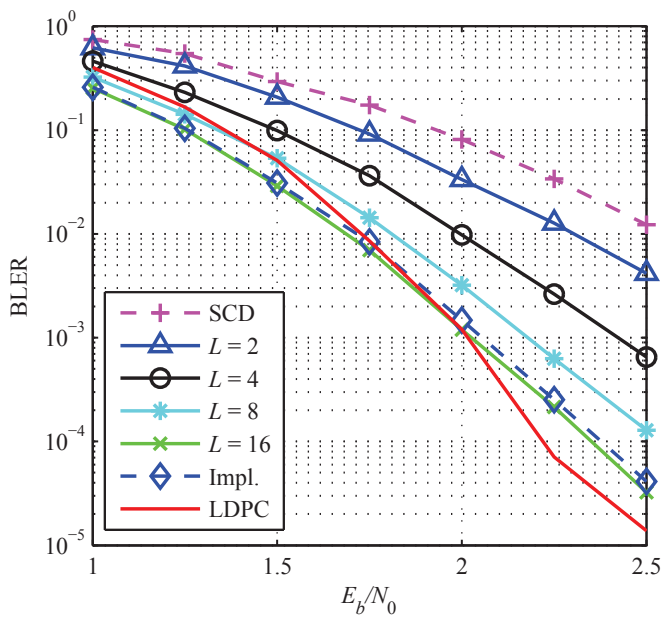}\caption{BLERs of LSCD with different list sizes.}
\end{figure}

\subsection{Implementation Results of the Low-latency LSCD}

The LSCD architecture proposed in Fig. 6 is designed and implemented
for an $\left(N,R,r\right)=\left(1024,1/2,16\right)$ polar code with
list size $\mathcal{L}=16$. $M=64$ PEs are used for each SCD. From
the simulation results, the SE method with $\epsilon=0.3$ and the
DTS-Advance with $RT=\gamma_{11}^{i}$ introduce negligible degradation
in the error-correcting performance, and hence they are used for the
hardware implementation. Fig. 13 compares our implementation's error-correcting
performance with those of the conventional LSCD with different list
sizes. It can be seen that our LSCD architecture has a very similar
BLER performance to the conventional LSCD. As a reference, the performances
of SCD and an $\left(N,R\right)=\left(1152,1/2\right)$ LDPC code
used in the WiMAX standard \cite{wimax} are also shown in Fig. 13.
Here, 40 iterations are used for the LDPC decoding. It can be seen
that polar codes have better performance when LSCD with a larger list
size $\mathcal{L}$ is used. When LSCD with $\mathcal{L}=16$ is used,
the BLER performance of polar codes is comparable to that of the LDPC
code. 

The design is synthesized with a UMC 90 nm CMOS process, using Synopsys
Design Compiler. For a fair comparison, the quantization scheme in
\cite{sTSP_EPFL} is used, i.e., the LLR and the path metric are represented
in 6 bits and 8 bits, respectively. Table V summarizes the synthesis
results and compares them with those of the existing architectures.
Compared with the state-of-the-art architectures, our proposed LSCD
architecture supports a much larger list size, which results in a
comparable error-correcting performance with other advanced error-correcting
codes. Moreover, from Table III, the proposed LSCD architecture requires
1462 clock cycles to decode one codeword, and hence it achieves a
decoding throughput of 460 Mbps at a clock frequency of 658 MHz. Compared
with \cite{ISCAS_Lehigh} and \cite{sTSP_EPFL}, both the decoding
throughput and the list size are doubled. The chip area presented
in Table V is mainly due to the state memory module. The SCD module
only occupies 0.53 $\textrm{mm}^{2}$ and the area of the LM module
is smaller than 0.1 $\textrm{mm}^{2}$.

\section{Conclusion}

In this work, a low-latency LSCD architecture is presented, which
is optimized at the system, algorithmic, and architectural levels.
At the system level, a selective expansion method is proposed such
that the amount of LM operations and the associated latency of the
reliable information bits are reduced. At the algorithmic level, a
double thresholding scheme is proposed as an approximate sorting method
for the list pruning operation and its logic delay is greatly reduced
for a large list size. Finally, an optimized VLSI architecture for
the LM operation is presented. Experimental results show that both
the decoding throughput and the list size are doubled when compared
with the state-of-the-art architectures.

\ifCLASSOPTIONcaptionsoff \newpage{}\fi



%\bibliographystyle{IEEEtran}

\begin{IEEEbiography}[{\includegraphics[width=1in,height=1.25in,clip,keepaspectratio]{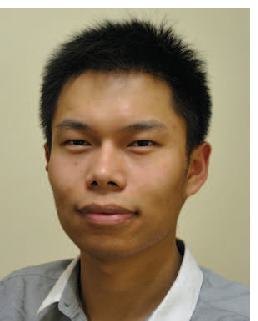}}]{YouZhe
Fan} (S'11-M'15) received the B.E. degree in electronic engineering
form the Harbin Institute of Technology, China, in 2009, and Ph.D.
degree in electronic and computer engineering from the Hong Kong University
of Science and Technology (HKUST), Hong Kong, in 2015, respectively.

He is now a Research Associate in the Department of Electronic and
Computer Engineering at the HKUST. His research interests are VLSI
architectures and integrated circuit design for communications and
coding theory applications, digital signal processing systems, and
general purpose computing systems. He is currently working on low-power
high-speed VLSI design for wideband wireless MIMO communications and
advanced error-control coding schemes such as low-density parity-check
(LDPC) codes and polar codes. \end{IEEEbiography}

\begin{IEEEbiography}[{\includegraphics[width=1in,height=1.25in,clip,keepaspectratio]{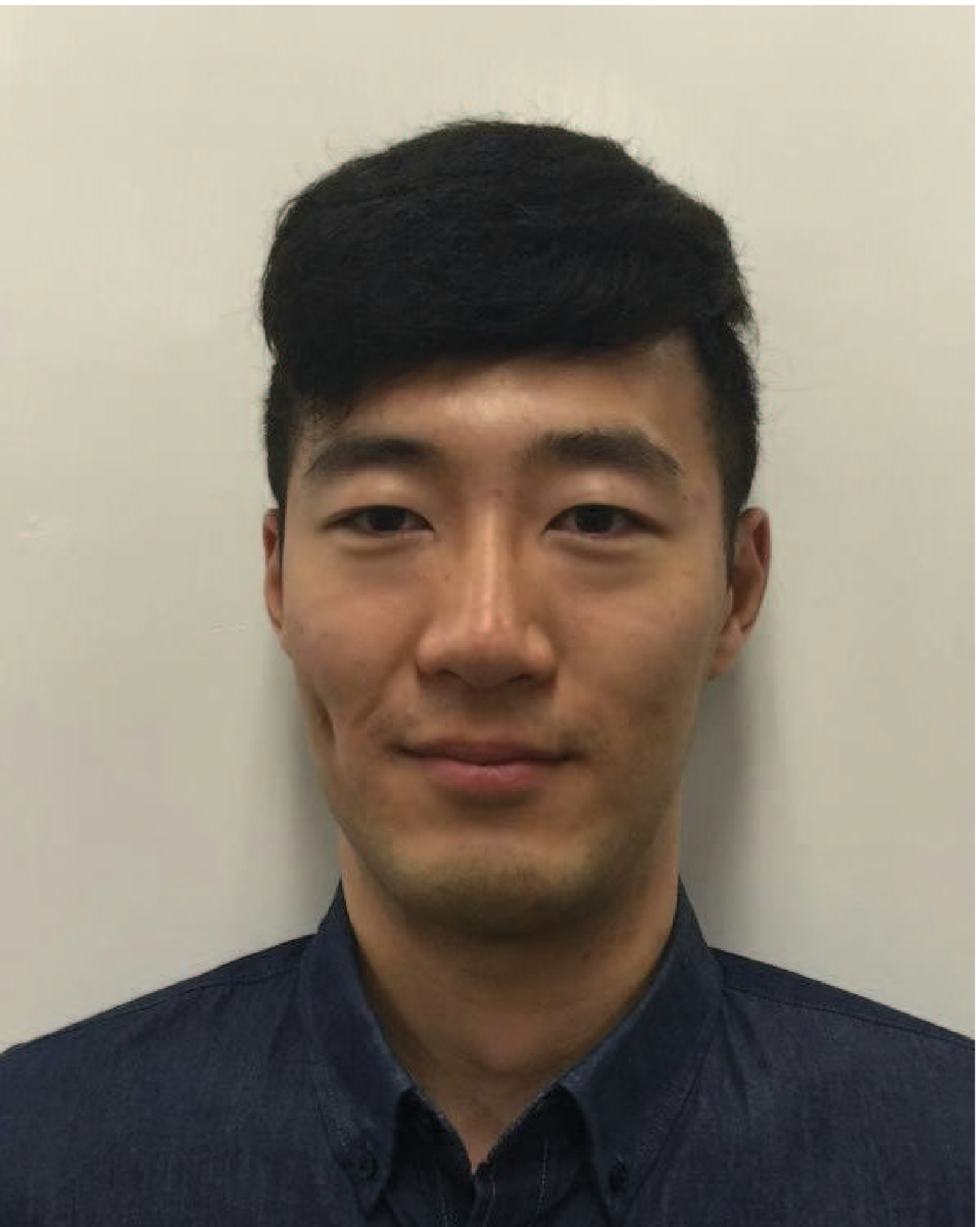}}]{ChenYang
Xia} (S'15) received the B.E. degree in electronic engineering from
Shanghai Jiao Tong University, China, in 2013. He is currently pursuing
the M.Phil. degree at the Department of Electronic and Computer Engineering,
the Hong Kong University of Science and Technology, Hong Kong.

His research interests include VLSI architecture and implementation
for communication systems and other digital signal processing systems.
He is currently working on high-speed low-complexity FPGA design for
channel codec system such as polar codes. \end{IEEEbiography}

\begin{IEEEbiography}[{\includegraphics[width=1in,height=1.25in,clip,keepaspectratio]{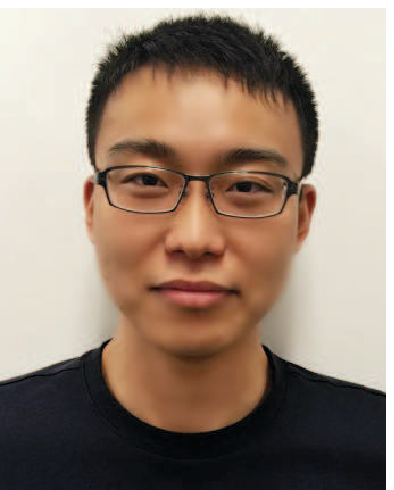}}]{Ji
Chen} (S'15) received the B.E. degree in electronic information and
communications from Huazhong University of Science and Technology
(HUST), China, in 2013. He is currently working towards the M.Phil.
degree in the Department of Electronic and Computer Engineering at
the Hong Kong University of Science and Technology (HKUST), Hong Kong.
His research interests are in information theory and signal processing.
He is currently working on the high-speed low-complexity decoding
algorithm design of polar codes. \end{IEEEbiography}

\begin{IEEEbiography}[{\includegraphics[width=1in,height=1.25in,clip,keepaspectratio]{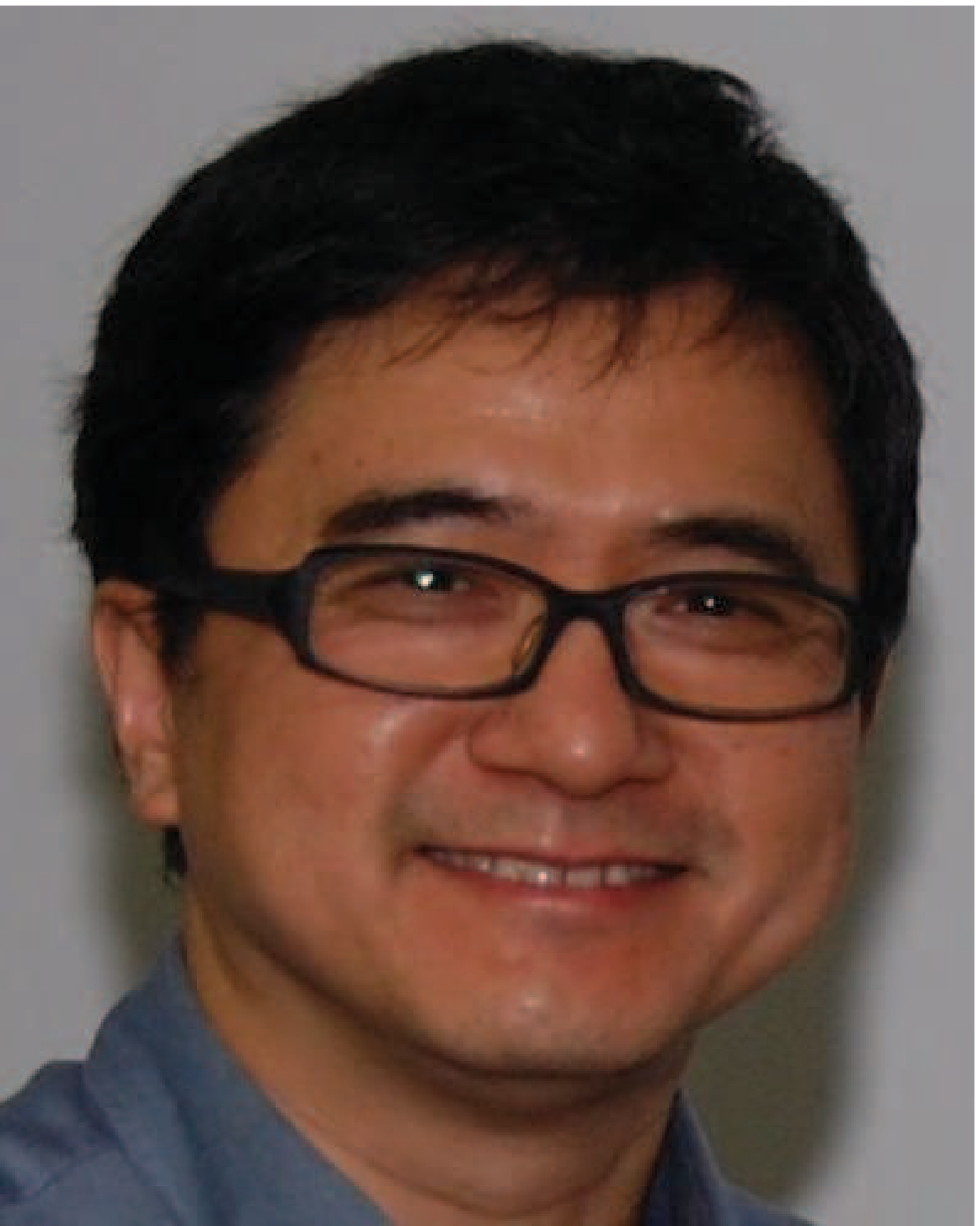}}]{Chi-ying
Tsui} (SM'11) received the B.S. degree in electrical engineering
from the University of Hong Kong and the Ph.D. degree in computer
engineering from the University of Southern California in 1994. 

He joined the Department of Electronic and Computer Engineering, Hong
Kong University of Science and Technology in 1994 and is currently
a full professor in the department. His research interests include
designing VLSI architectures for low power multimedia and wireless
applications, developing power management circuits and techniques
for embedded portable devices and ultralow power systems. He has published
more than 170 referred publications and holds 10 US patents on power
management, VLSI and multimedia systems. 

Dr. Tsui received the Best Paper awards from the \textsc{IEEE Transactions
on VLSI Systems} in 1995, IEEE ISCAS in 1999, IEEE/ACM ISLPED in 2007,
and IEEE DELTA in 2008, CODES in 2012. He also received the Design
Awards in the IEEE ASP-DAC University Design Contest in 2004 and 2006.
\end{IEEEbiography}

\begin{IEEEbiography}[{\includegraphics[width=1in,height=1.25in,clip,keepaspectratio]{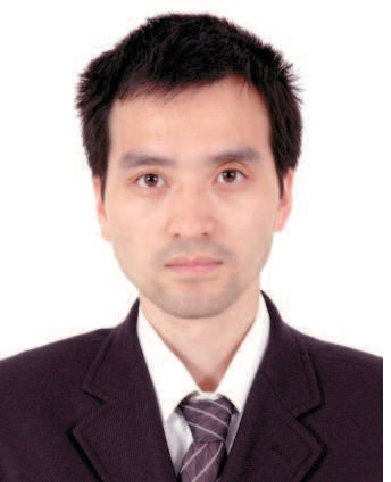}}]{Jie
Jin} received the B.S. degree in electronic engineering from Xi’an
Jiaotong University and Ph.D. degree in electronic and computer engineering
from the Hong Kong University of Science and Technology in 2009. He
joined Huawei Technologies in 2009 and is currently a senior research
engineer. His research interests include VLSI architectures for low
power communications and channel coding applications, and digital
signal processing systems. He is currently working on VLSI architectures
for advanced channel coding schemes such as low-density parity-check
codes and polar codes.\end{IEEEbiography}

\begin{IEEEbiography}[{\includegraphics[width=1in,height=1.25in,clip,keepaspectratio]{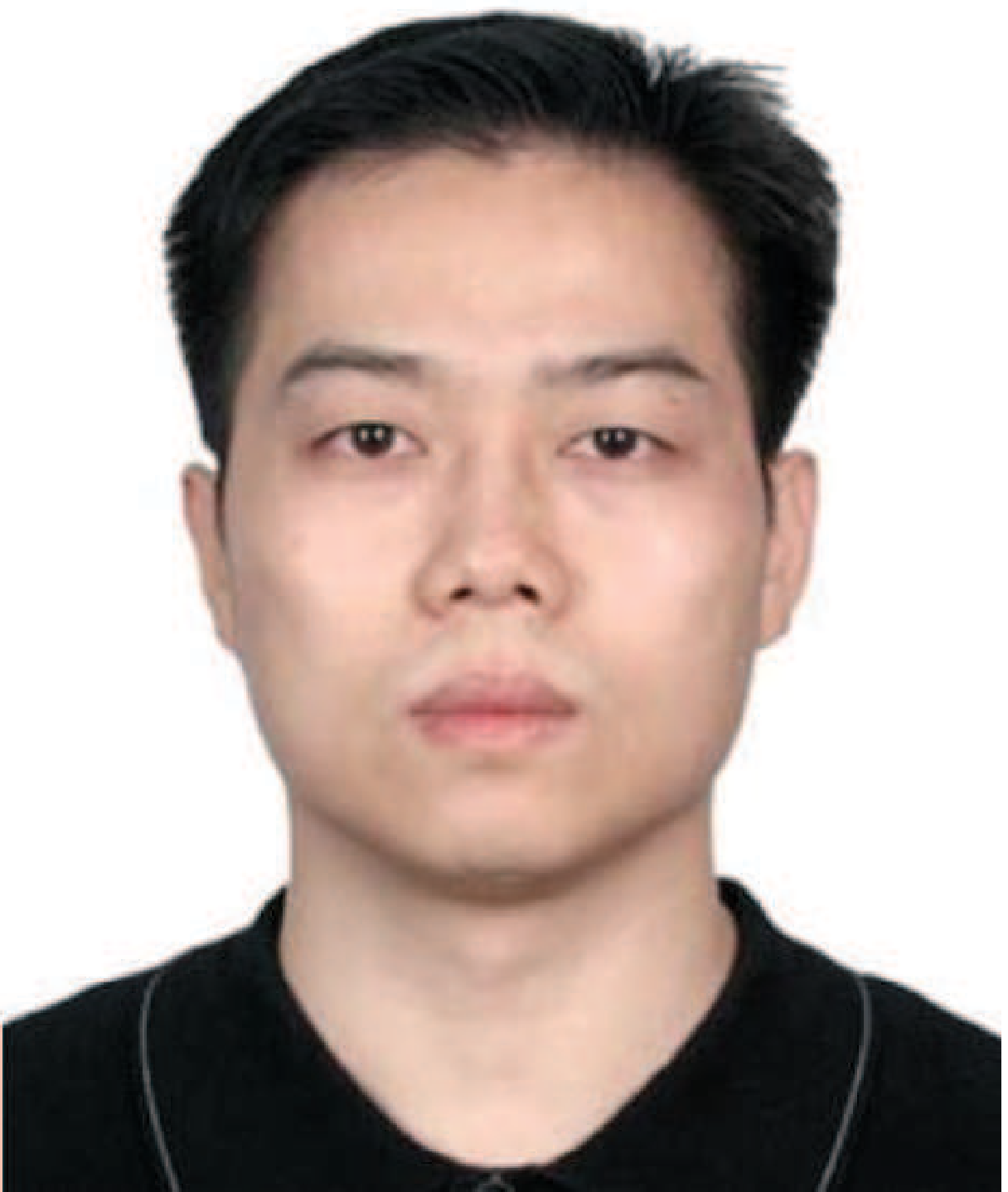}}]{Hui
Shen} (M'09) was born in 1975. He received the Ph.D. degree in electronics
and communication engineering from the Huazhong University of Science
and Technology, P.R.China in April 2004. From April 2004 to September
2007, he was with Technical Center, Research Department of ZTE Corporation,
Shenzhen, P.R.China as a researcher and standard senior engineer.
Currently, he is with Huawei Corporation, Shenzhen, P.R.China. His
research interests lie in the areas of wireless communications, design
and analysis of multiple-antenna systems, multi-user MIMO pre-coding,
interference alignment. \end{IEEEbiography}

\begin{IEEEbiography}[{\includegraphics[width=1in,height=1.25in,clip,keepaspectratio]{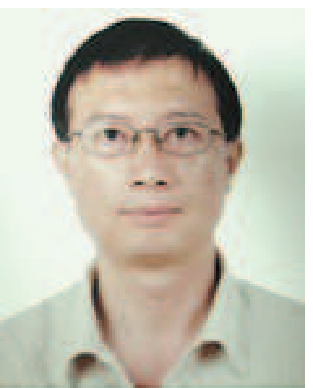}}]{Bin
Li} (M'08) received the Ph.D. degree in communications engineering
from the Nanjing Institute of Communications Engineering, Nanjing,
China, in 1993. From 1996 to 1997, he was a visiting professor with
the School of Engineering Science, Simon Fraser University, Canada.
From 1997 to 2001, he was a member of technical staff in Nortel, Ottawa.
From 2001 to 2005, he was a senior staff engineer in InterDigital,
NY, USA. Since November 2005, he has been a senior expert in Huawei
Technologies, Shenzhen, China. His research interests are modulation,
coding and MIMO. \end{IEEEbiography}

\vfill
\end{document}